# Modeling CSFs of B2C E-commerce Systems Using the Enterprise Architecture Approach

A thesis submitted as partial fulfillment of the requirements for the degree
of Master of Information and Communication Technology

Rayed AlGhamdi

MICT, Griffith University

Supervisor

Associate Professor Peter Bernus

School of Information and Communication Technology

Griffith University

Brisbane, Queensland, Australia

June 2008

## Declaration

The work in this thesis has not been submitted previously, in whole or in part, for any other academic award and is my original work, except where acknowledged. The work has been carried out since the beginning of my research program in August 2007.

Rayed AlGhamdi

June 2008

## Acknowledgement

I take this opportunity to acknowledge, appreciate and thank all those who have helped, supported and encouraged me throughout this research process. To my supervisor, Associate Professor Peter Bernus, many thanks for his kind and wise advice throughout this research. He provided extremely valuable assistance in all areas of the dissertation. To the Higher Education Ministry of the Kingdom of Saudi Arabia, in particular King Abdulaziz University, many thanks for offering me this opportunity (by providing a scholarship) to complete this degree and also for their generous financial support for me and my family. To my parents, special thanks for their devotion to me throughout my life journey, May God save them and give them good health. To my wife, Reem, and my daughter, Reena, many thanks for their unfailing love, support and their confidence in me during the time I have worked on this dissertation. Lastly, I wish to thank my brothers and sisters for their continual encouragement and support.

# Abstract

This study is to investigate the Critical Success Factors (CSFs) of the Business to Customer (B2C) e-commerce system. These factors should be considered comprehensively and expanded to all parties concerned to create and provide the electronic service and ensure that the CSFs are satisfied. In order to give an organized and inclusive view of the CSFs, an enterprise architecture framework will be adopted to systemize this investigation.

The adoption of the enterprise architecture framework gives several benefits to this project:

- It gives a complete view of the B2C e-commerce organization in order to consider all involved activities in providing eService,
- It shows all involved entities in providing eService and their relationships,
- It expresses the CSFs as principles as part of the 'concept' of the involved entities; and therefore,
- It organizes and structures what needs to be done in the enterprise and its entities in order to satisfy the CSFs.

In the literature reviewed, there are fairly extensive studies that have proposed a range of significant factors that could be considered as factors for the successful operation of B2C e-commerce systems. However, not many studies have evaluated the factors for *building* successful B2C e-commerce systems. For this reason, an enterprise architecture framework was adopted, because such frameworks can represent the life cycle relationships between entities such as the entity that provides the B2C service, and the entity (or entities) that create such a service. This adoption helps to consider various aspects throughout building and launching the B2C eService system. This consideration helps to ensure that the CSFs are satisfied.

Literature available on factors contributing directly and indirectly to the success of B2C e-commerce systems have been analyzed, and a total of 29 B2C e-commerce success factors have been identified. These factors are divided into two groups; the first having a direct relationship to the success of the B2C eService, and the second contributing indirectly.

The identified critical success factors are mapped on a typical model of B2C business. This mapping is intended to:

(1) Demonstrate the responsibilities of the different entities and their relationships in an enterprise in order to satisfy the CSFs;

(2) Classify the CSFs in new structure to be able to concentrate on the underlying system properties, from which the rest of the properties emerge; and

(3) Help to ensure that the CSFs are satisfied by designing a process and identifying the different responsibilities in that process for the different areas of the enterprise.

The new classification structure of the CSFs illustrates the hierarchal structure of the CSFs. This structure clarifies the underlining critical success components. B2C e-commerce success is influenced by two main factors: customer satisfaction and loyalty, and enabling enterprise processes. Furthermore, both customer satisfaction and loyalty and enterprise processes are influenced by a number of factors.

In order to discuss what needs to be done to ensure that the CSFs are satisfied, a process model of B2C eService development was established, concentrating on the necessary activities, their outcomes and input/output relationships, as well as identifying the role of Enterprise Business Entities (EBEs) in this process. Through a better understanding of the B2C eService development process, the CSFs of the eService development project itself can be comprehended, as well as its relationship to the CSFs of the B2C Service. In order to analyze and understand this process, an activity modeling language (IDEFØ) was used.

Due to the limited scope of this thesis, one example CSF is chosen, namely 'ease of use' (or in general 'usability'), as a critical success factor of the B2C e-commerce system to be discussed. All other CSFs could be analyzed in a similar manner, and the developed life cycle process model of the involved entities could be extended with the necessary activities to satisfy these CSFs.

In order to ensure that the CSFs are satisfied, one must consider:

- *The quality of the B2C system itself.*
- *The quality of the eService development project.*
- *The quality of the B2C system's management process.*

Finally, as far as this thesis is concerned, the validity of the results hinges on correct interpretations of the CSFs reported in the literature; therefore, it is expected that future research needs to be done after this thesis is complete to confirm this interpretation. Furthermore, this research could be further extended using other methods that can add value to this study. One possibility is to do a case study with observation of the use of these results to provide feedback on the research. Another method is to conduct action research, which allows practice in the development of the B2C system. This way of conducting research adds value to improving the results instead of knowing the validity/ usability of the research outcomes. Moreover, the business model used in this study can be enriched by mapping other CSFs onto it in a similar manner by:

- adding the relevant information (such as standards) to the developed life cycle process model, or
- adding necessary new activities to the process model

to satisfy these CSFs.

# Table of Contents

## List of Figures

# 1. Introduction

The number of commercial organizations that tend to apply electronic commerce systems is on the rise. In the near future, this trend will become not only a tool to simply increase income, but without doubt will be considered an essential means to compete with other companies. From this perspective, the application of electronic commerce systems without a study of the success factors in this area may lead to a failure in practice.

This study is to investigate the Critical Success Factors (CSFs) of Business to Customer (B2C) e-commerce systems. These factors should be considered comprehensively and expanded to all parties concerned to provide the electronic service. In order to give an organized and inclusive view of CSFs, a framework will be adapted to systemize this investigation. The enterprise architecture approach provides a solution to express CSFs as principles as part of the 'concept' of involved entities.

The study supposes that there are three options for building a B2C online market for a commercial organization: (1) an organization knows how to initiate eService and has the essential skills to implement it (i.e. this does not exclude the use of external contractors); (2) an organization knows they want to create eService, but they need to get prepared to learn what and how can be done; (3) an organization knows they want to create eService and will let another organization to do it for them. However, the last option is not recommended for the following two reasons:

a) The new service will be a 'foreign body' to the organization, without the skills and knowledge of how the service can be maintained and continuously improved. Thus, there is a minimum level of competency needed.

b) It makes the organization fully dependent on the external service to operate and maintain the new service.

Thus, the focus of this study will be on the first option: the company knows how to initiate eService and has the skills to implement it. Yet, the second option will be mentioned in the section on future research and a suggested business model will be proposed.

## 1.1. Research questions

The main question in this study is:

What are the CSFs in creating and operating B2C systems (enterprises)?

Furthermore, there are two sub-questions:

(1) What are the CSFs for e-service development projects and the CSFs for operating and managing B2C systems?

(2) What needs to be done by the enterprise to ensure that the CSFs are satisfied?

## 1.2. Research Aims and objectives

This study aims to:

- Identify the CSFs of B2C e-commerce systems in the literature.
- Develop the scope using a business model of enterprise architecture.
- Map the identified CSFs onto the business model in order to demonstrate the responsibilities of the different entities and their relationships in an enterprise.
- Classify CSFs in a new structure to be able to concentrate on the underlying system properties from which the rest of the properties emerge.
- Identify new areas (potential CSFs).

## 1.3. Research Significance

According to the proposed objectives of the thesis, the following contributions are expected:

The research (1) structures B2C e-commerce systems, including the representation of the structures needed for their creation, (2) then systematizes findings of the literature about CSFs, and (3) through this systematic approach identifies potential new areas that may have to be considered, e.g. 'Preparedness Projects' - where there may be CSFs not previously studied.

## 1.5. List of Abbreviations

| | |
|---|---|
| B2C | Business to Customer |
| CES | Customer E-commerce Satisfaction |
| CRM | Customers Relationship Management |
| CSFs | Critical Success Factors |
| EBEs | Enterprise Business Entities |
| EC | Electronic Commerce |
| ECCRM | Electronic Commerce Customers Relationship Management |
| EDP | E-service Development Project |
| ERP | Enterprise Resource Planning |
| eService | Electronic Service |
| GERA | Generalised Enterprise Reference Architecture |
| GERAM | Generalised Enterprise Reference Architecture and Methodology |
| HCI | Human-Computer Interaction |
| HQ | Headquarters |
| IDEF Ø | Integrated DEFinition methods |
| IS | Information System |

## 1.5. Research Organization (Structure)

This study includes five chapters. The first chapter gives a brief introduction to this study. Through a critical review of the literature, the second chapter identifies involved factors of success in providing eService. The third chapter explains the research framework and a systematic framework for CSFs. In the fourth chapter, the identified factors will be organized by adapting a framework that will systemize this investigation in order to give an organized view of CSFs. It investigates the structure of the business transformation through applying the framework to the selected business problem, and the identified CSFs are mapped on the selected business model. This mapping intends to:

(1) Demonstrate the responsibilities of the different entities and their relationships in an enterprise in order to satisfy CSFs;

(2) Classify CSFs into a new structure to be able to concentrate on the underlying system properties from which the rest of the properties emerge; and

(3) Help to ensure that the CSFs are satisfied by designing a process and identifying the different responsibilities in that process for the different areas of the enterprise. Finally, the study's conclusion and future research are discussed in the fifth chapter.

## 2. Involved Factors of Success in Providing eService

Extensive studies have proposed a range of significant factors that could be considered for the successful operation of B2C e-commerce systems. However, not many studies have evaluated the factors for *building* successful B2C e-commerce systems.

### 2.1. Factors of success in the literature

Through a literature review and interviews with managers in e-commerce companies, Sung (2006) listed 16 critical success factors of e-commerce:

- customer relationship,
- customers' information privacy,
- low-cost operation of the e-commerce system,
- ease of using an e-commerce website,
- e-commerce strategy,
- technical e-commerce expertise,
- stability of the system,
- security of the system,
- speed of the system,
- abundant information about goods and services provided to customers,
- variety of goods/services,
- variety and safety of payment processes,
- customer services and support,
- the accuracy of goods/services delivery,
- low price of goods and services (competitive), and
- continuous evaluation of e-commerce operations.

The study was conducted in three countries: USA, Korea and Japan. In the three countries, the customer relationship, ease of use, variety of goods/services and the accuracy of delivery of goods/services were found to be the most important e-commerce CSFs.

Holsapple and Sasidharan (2005) focused on trust, as it is relevant to Business to Consumer (B2C) e-commerce transactions. They identified that *trust* is a fundamental key that determines the success of B2C e-commerce processes.

Through a case study method of investigation and the experiences of six online companies, Dubelaar, Sohal and Savic (2005) outlined six major CSFs in B2C e-business adoption:

- combining e-business knowledge and value proposition,
- replication of offline brand,
- building trust,
-  measuring performance and value delivery,
- customer satisfaction and retention, and finally
- monitoring internal processes and competitor activity.

Hassanein and Head (2004) ascertained that online *trust* is a critical factor for e-commerce transactions and is fundamental in B2C e-commerce systems. They emphasize the importance of building online trust through a '*socially rich*' web interface. Their investigation reaches the conclusion that online trust can be built by infusing social presence on a commercial website through text descriptions and pictures that speak to the customer's emotions. Moreover, besides the *social presence* factor, the research model contains other three factors that can influence online trust:

- perceived usefulness,
- perceived ease of use, and
- enjoyment.

In another study, Gefen and Straub (2004) stated that *trust* is a central element in many commercial processes. They validate "*a four-dimensional scale of trust in the context of e-Products and revalidates it in the context of e-Services. The study then shows the influence of social presence on these dimensions of this trust, especially benevolence,* *and its ultimate contribution to online purchase intentions*" (Gefen & Straub 2004, p.407).

Molla and Licker (2001) extended the Delone and Maclean Model of IS (Information System) success to e-commerce systems. They proposed that Customer E-commerce Satisfaction (CES) is a dependent factor to e-commerce success. It has critical relationships with five variables (see Fig 1):

- e-commerce system quality,
- content quality,
- ease of use,
- trust, and
- support and service.

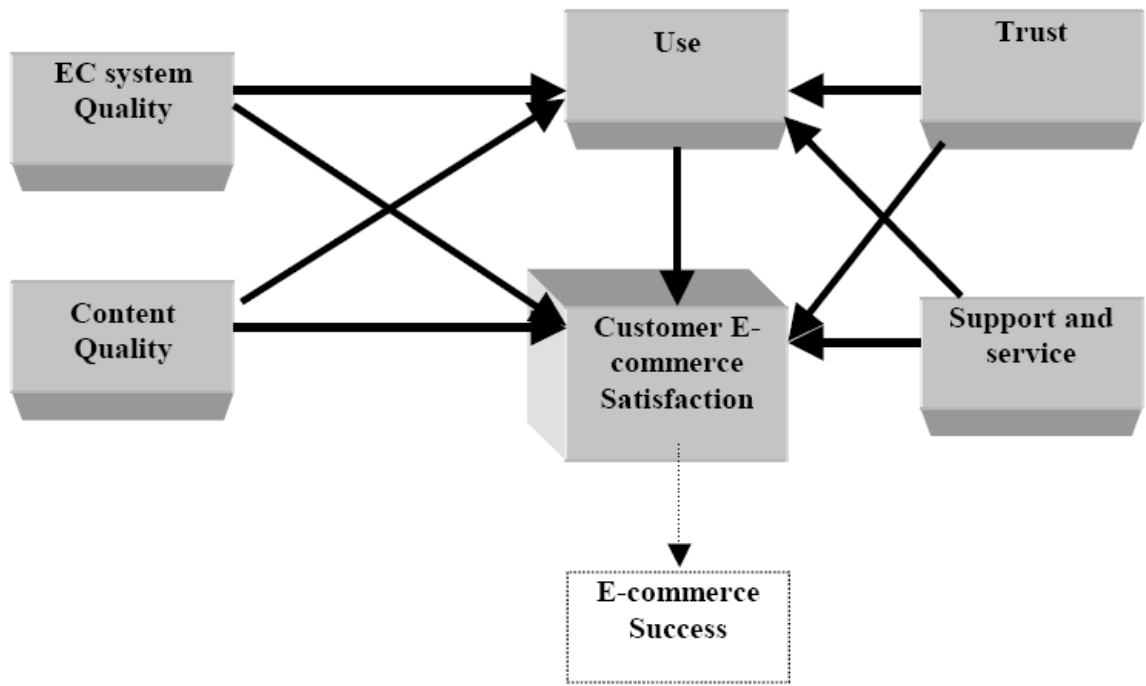

Fig. 1: E-commerce Success Model
Source: Molla & Licker 2001, p.136

Ribbink et al. (2004) demonstrated that customer loyalty is one of the most important factors for B2C success. Furthermore, they show that customer loyalty is a result of customer satisfaction and trust and those two factors are obtained by five other factors:

- ease of use,
- website design,

- customization,

- responsiveness, and

- quality assurance.

Terry and Standing (2004) mentioned that *user participation* in a B2C e-commerce system development is essential to the success of the system. As a consequence, user participation has a direct positive effect on user satisfaction and is therefore another factor for a system's success. Naturally, companies need to devise means of involving end users (e.g. focus groups, evaluations and testing instruments) to satisfy this requirement.

Wirtz and Lihotzky (2003) noted that customer retention in B2C e-business systems is particularly important. Therefore, they investigated how customer retention can be accomplished in B2C electronic business. Their investigation showed that there are seven related retention strategies used extensively and successfully by B2C e-business systems at present:

- trust building,

- community,

- convenience,

- free services,

- individualization,

- contractual agreements, and

- technical integration.

However, not all the previous strategies are suitable for all of the e-business models. They divide the e-business models on the Internet into four models: content, commerce, context, and connection. The commerce model is relevant to this study. Thus, there are three strategies that have the most influence on the B2C e-commerce model:

- trust building,

- convenience, and

- individualization.

Individualization and convenience lead to increased customer satisfaction, and trust leads to building strong relationships and commitment.

Chen, Chen and Kazman (2007) concentrated on websites that work as online markets and that can create effective connection channels between agents (producers) and customers (consumers) through creating excellent design, diverse alternatives and touchable dealings. They informed that: "*Touch is a multidimensional concept which encompasses three aspects: (1) communication media characteristics (e.g., media richness), (2) the affective quality of touch options (e.g. humanization) and (3) interactivity between the online customers and the system*" (Chen et al., 2007, p.73).

Schodeler and Madeja (2004) found that Electronic Commerce Customer Relationship Management (*ECCRM*) has a particularly strong influence on the success of B2C companies.

Green and Pearson (2006) paid extreme attention to *website usability* as a critical success factor of the B2C e-commerce system. They stated that poor website design is one of the recent critical failure factors for B2C e-commerce systems.

Ryan and Valverde (2006) reported that the quickness of handling e-service gives a good impression to online customers, which leads to customer satisfaction. In contrast, delays and waiting on the internet are a contributing factor in e-commerce failures.

Song and Zahedi (2001) found that *effective website design* plays a critical role in attracting online customers and more frequent purchases. Furthermore, efficient website design can help an organization to build good relationships with its customers. Song and Zadehi (2001, p.207) developed a model "*for the process by which web design elements could influence the purchase intention of online customers. Website design elements are defined as the features, components and information used in developing e-commerce website which may influence customers' purchase behavior through the reinforcement of their positive beliefs*". They define various website design elements that refer to five categories: promotion, service, information influence, self-efficacy, and resource facilitation. Examples of these elements are: price discounting presentation, feedback section, customer rating of product, currency conversion option, frequently asked questions, and various shipping and payment options.

Based on a theoretical model, Atchariyachanvanich, Okada and Sonehara (2007) distinguished five factors that make online customers repurchase through an online market: satisfaction, confirmation, perceived usefulness, customer loyalty, and perceived incentives. Maintaining online customers' repurchase through an online market means achieving success.

Kim and Lee (2002) have identified critical design factors that affect the performance of e-commerce systems. They classified design factors into two groups: the *process* and the *architecture* perspectives, which are defined as follows "*The process perspective categorizes design factors into four groups according to the transaction phases: information, agreement, settlement and environment. The architecture perspective categorizes design factors into four groups according to the web architecture, which are content, structure, interaction and presentation*" (p.187). The process perspective is generally more extensive than the architecture perspective, because the former includes the entire transaction process, whereas the latter focuses on the system implementation details. These design factors contribute towards the design and development of an effective e-commerce system which is easy to use for online customers.

Through a conceptual framework, Shankar, Smith and Rangaswamy (2003) developed "*hypotheses about the effects of the online medium on customer satisfaction and loyalty and on the relationships between satisfaction and loyalty*" (p.153). They found that in an online market environment, customers' "*loyalty has a strong positive reciprocal impact on overall satisfaction*" (p.172) and "*that overall satisfaction had stronger positive impact on loyalty*" (p.172). Furthermore, the "*ease of obtaining information has a stronger positive effect on overall satisfaction*" (p.172).

Udo and Marquis (2001) identified four factors which are critical for e-commerce website effectiveness. These factors are: download time, ease of navigation, appropriate use of graphics and interactivity.

 Santos (2003) stated the eService quality is a key determinant for e-commerce success. Based on a user focus group study, the researcher proposed a "*conceptual model of the determinants of the eService quality*" (p.133). E-service quality is divided into two groups of factors (before and after the launching of eService website): incubative and active dimensions. The incubative dimension is during the design phase which includes ease of use, appearance, linkage, structure and layout, and content. The active dimension is needed when the website is established "*to be maintained continuously*" (p.237). It includes reliability, efficiency, support, communication, security, and incentives.

In summary, the literature review shows a wide range of issues that relate to the management of customer relationships, e-commerce system efficiency, product characteristics and costs, e-business strategies, website usability/ease of use, security and privacy, support and services, and resource management. The following section (section 2.2) charts CSFs identified in the literature and indicates their relationships contributing to B2C e-commerce success.

## 2.2. Identified factors of success in the literature

Literature available on factors contributing directly and indirectly to the success of B2C e-commerce systems have been analyzed and a total of 29 B2C e-commerce success factors have been identified. These factors are divided into two groups; the first having a direct relationship to the success of the B2C eService, and the second contributing indirectly (see Fig. 2).

### 2.2.1. Relationships among Identified B2C e-commerce critical success factors (explanation of Figure 2)

The reviewed sources named 15 critical success factors for B2C e-commerce systems that have been found to contribute directly to B2C e-commerce success. Customer satisfaction and trust are central to all of these factors. The remaining factors contribute to B2C e-commerce success through their relationships with these central two factors. These factors are responsiveness/interaction, convenience, support and services, customization, content quality, e-commerce system quality, speed of system, website design, ease of use, usefulness, system security, social presence and information privacy. Furthermore, customer loyalty contributes directly to B2C e-commerce success and is achievable over time through building trust and ensuring customer satisfaction (Yeo & Chiam, 2006; Ribbink et al., 2004).

The reviewed sources named other 14 critical success factors and their indirect bearing on the success of the B2C system. The CSFs in question are: payment processes, e-commerce strategies, Electronic Commerce Customer Relationship Management (ECCRM), low cost operation, user participation in the development process, low price of products/services, variety of goods/services offered, quality of goods/services delivery, evaluation e-commerce operation, product quality, replication of offline brand, measuring performance and value delivery, technical e-commerce expertise, combining e-business and value proposition, and monitoring internal processes and competitor activity. Even though the direct effects of these have not been established by the literature, they still contribute to the B2C e-commerce system's viability.

The above direct and indirect relationships are charted in Fig. 2. Further discussion of the identified CSFs will be provided later in section 4.2.

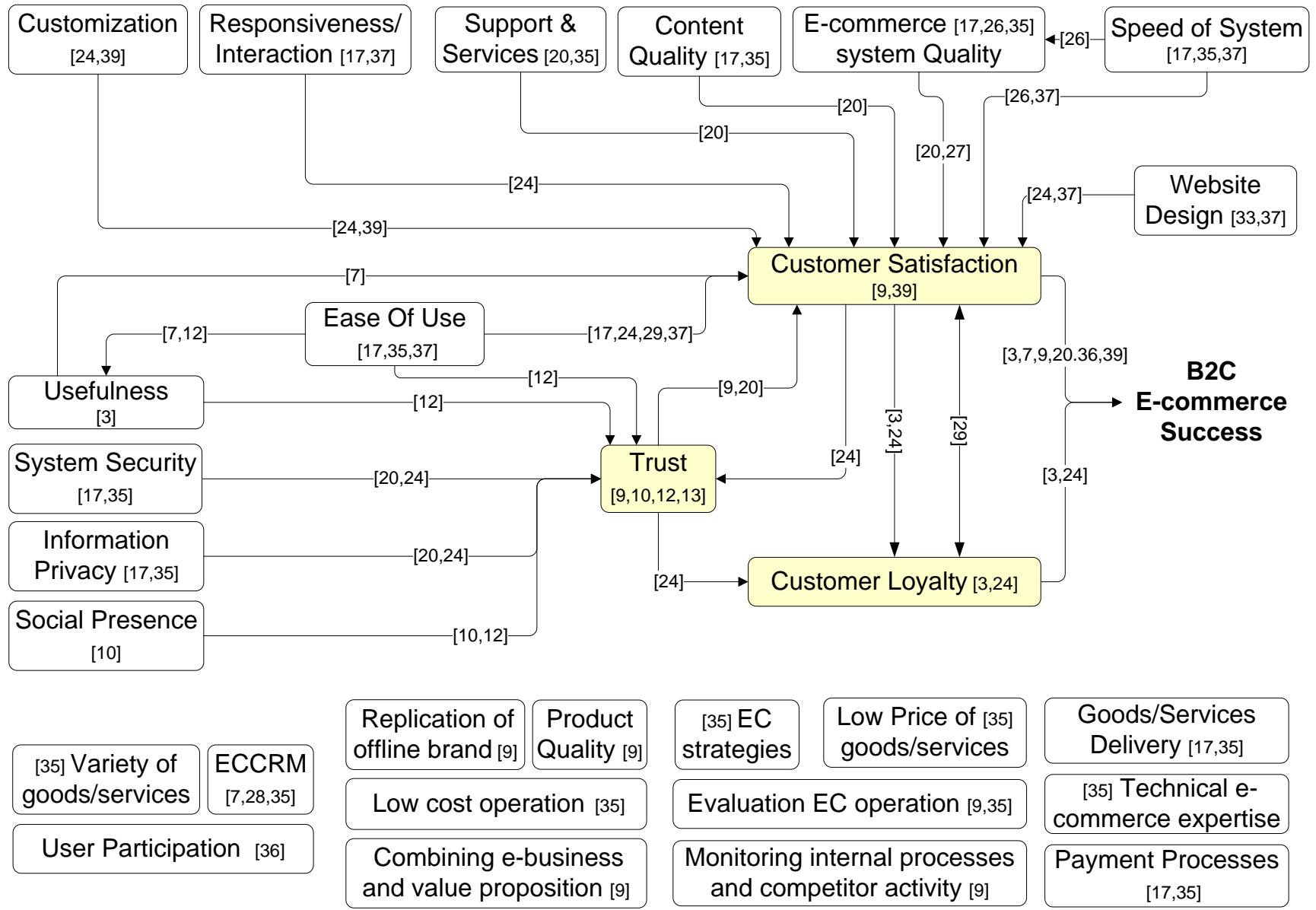

Fig. 2: Relationship Chart of the Identified B2C e-commerce critical success factors

### 2.2.2. Explanation of the Flow Chart (Fig. 2 notation)

The Flow Chart consists of boxes, arrows and numbers which are shown inside the boxes and on the arrows. A box represents a critical success factor that influences the success of the B2C e-commerce system. The arrows represent the relationship between one factor and another. The number inside a box refers to the literature that identifies the factor as a critical success factor for B2C e-commerce systems. The numbers on the arrows refer to the literature that explains the relationship between two factors.

### 2.2.3. Referencing of the identified CSFs

- Combining e-business and value proposition (Dubelaar, Sohal & Savic, 2005)[9]
- Content Quality   (Kim & Lee, 2002)[17], and (Sung, 2006)[35]
- Customer loyalty   (Atchariyachanvanich, Okada & Sonehara, 2007)[3], and (Ribbink et al., 2004)[24]
- Customer satisfaction (Dubelaar, Sohal & Savic, 2005)[9], and (Wirtz & Lihotzky, 2003)[39]
- Customization (Ribbink et al., 2004)[24], and (Wirtz & Lihotzky, 2003)[39]
- Ease of use   (Kim & Lee, 2002)[17], (Sung, 2006)[35], and (Udo & Marquis, 2001)[37]
- EC strategies (Sung, 2006)[35]
- E-commerce system quality (Kim & Lee, 2002)[17], (Santos, 2003)[27], and (Sung, 2006)[35]
- Electronic Commerce Customers Relationship Management (ECCRM) (Chen, Chen & Kazman, 2007)[7], (Schodeler et al., 2006)[28] and (Sung, 2006)[35]
- Evaluation E-Commerce (EC) operation  (Dubelaar, Sohal & Savic, 2005)[9] and (Sung, 2006)[35]
- Goods/Services delivery (Kim & Lee, 2002)[17], and (Sung, 2006)[35]
- Information privacy  (Kim & Lee, 2002)[17], and (Sung, 2006)[35]
- Low cost operation (Sung, 2006)[35]
- Low price of goods/services  (Sung, 2006)[35]

- Monitoring internal processes and competitor activity (Dubelaar, Sohal & Savic, 2005)[9]

- Payment processes (Kim & Lee, 2002)[17] and (Sung, 2006)[35]

- Product Quality (Dubelaar, Sohal & Savic, 2005)[9]

- Replication of offline brand (Dubelaar, Sohal & Savic, 2005)[9]

- Responsiveness and interaction (Kim & Lee, 2002)[17] and (Udo & Marquis, 2001)[37]

- Social presence (Gefen & Straub, 2004)[10] and (Hassanein & Head, 2004)[12]

- Speed of system (Kim & Lee, 2002)[17], (Sung, 2006)[35], (Udo & Marquis, 2001)[37]

- Support and Services (Molla & Licker, 2001)[20], and (Sung, 2006) [35]

- System security (Kim & Lee, 2002)[17], and (Sung, 2006)[35]

- Technical e-commerce expertise (Sung, 2006)[35]

- Trust (Dubelaar, Sohal & Savic, 2005)[9], (Gefen & Straub, 2004)[10], (Hassanein & Head, 2004)[12], and (Holsapple & Sasidharan, 2005)[13]

- Usefulness (Atchariuachanvanich et al., 2006)[3]

- User participation (Terry & Standing, 2004)[36]

- Variety of goods/services (Sung, 2006)[35]

- Website design (Song & Zahedi, 2001)[33], and (Udo & Marquis, 2001)[37]

# 3. Research Methodology: a Systematic Framework for CSFs

## 3.1 Research Methodology

This research concentrates on the critical success factors of B2C e-commerce systems using an enterprise architecture approach. The literature review is performed to identify the CSFs of B2C e-commerce systems that have been identified by other researchers. The literature review continues throughout the whole study, because during the analysis phase, new areas may become apparent which need to be investigated. After identifying the CSFs, a business model is developed to understand the scope of the CSFs. The identified CSFs are mapped onto the business model. This mapping helps to identify and describe all involved activities in order to satisfy the CSFs. Furthermore, this mapping can identify new areas (CSFs). In case of identifying new areas, a revision of the literature needs to be made to ensure that the identified new areas have not been discussed in the literature. The outcome of this thesis is to (1) systematize the CSFs, and (2) identify new areas (potential CSFs). See Fig. 3 below.

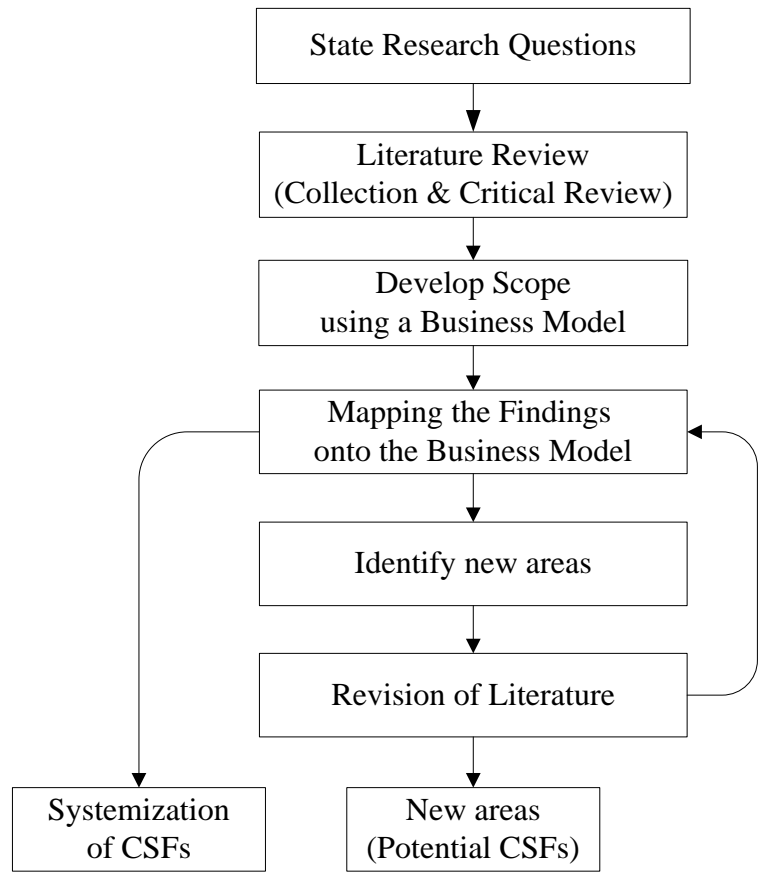

Fig. 3: Research Methodology

Data gathering and analysis techniques in this study include the use of the research framework to develop hypotheses and the analysis of research literature. Questionnaire and interview methods have been considered, but rejected, because it can be expected that the critical information required to answer the research questions is part of a successful organization's competitive strategy, and therefore would not be freely available (it is confidential). Although a solution may be possible to overcome this hurdle; the scope (time and resource limitation) of this project does not allow these methods to be used.

In order to give an organized and inclusive view of the CSFs, a framework will be adopted to systemize this investigation. The enterprise architecture approach provides a solution to express the CSFs as design principles (as part of the 'concept' guiding the specification and design of the involved entities).

Thus, as explained at the beginning of chapter 2 regarding the available literature, fairly extensive studies have proposed a range of significant factors that could be considered for the successful operation of B2C e-commerce systems. However, not many studies have evaluated the factors for *building* successful B2C e-commerce systems. For this reason, an enterprise architecture framework is adopted, because such frameworks can represent the life cycle relationships between entities, such as the entity that provides the B2C service, and the entity (or entities) that create such a service. This adoption helps to consider various aspects throughout building and launching the B2C eService system. This consideration helps to ensure that the CSFs are satisfied.

The adoption of the enterprise architecture framework gives several benefits to this project: it gives a complete view of the B2C e-commerce organization in order to consider all involved activities in providing eService; it shows all involved entities in providing eService and their relationships; it expresses the CSFs as principles as part of the 'concept' of involved entities; and therefore, it organizes and structures what needs to be done in the enterprise and its entities in order to satisfy the CSFs.

This study is conducted under the constructive positivism approach, because it is assumed that the results will be an objective and the correct interpretations of the CSFs as well as the relationships of these belong to the process to create a B2C system.

Useful results are expected; however, further work is needed. This research could be further extended using other methods that can add value to this study. One possibility is to do a case study with observation of the use of these results, which would provide feedback to the research. Another method is to conduct action research which allows practice in the development of the B2C system. This way of conducting research adds value to improve the results instead of simply knowing the validity/usability of the research outcomes.

As far as this thesis is concerned, the validity of the results depends on correct interpretation of the CSFs reported in the literature; therefore, it is expected that future research needs to be done after this thesis is completed to confirm this interpretation.

## 3.2 Research Framework

In this study, it is necessary to provide a comprehensive and in-depth view of the CSFs in the construction, development and operation of a B2C eService. The enterprise architecture approach gives a *complete* description of the involved processes in the enterprise in order to make sure those CSFs are satisfied (IFIP-IFAC Task Force, 2003). Throughout the complete life cycle of the involved entities that provide eService, validation and verification are necessary to maintain these CSFs in each phase.

In order to achieve this goal, the GERAM (Generalised Enterprise Reference Architecture and Methodology) is a candidate, because it is able to define the scope of activities in any business and therefore can categorize the CSFs. Since the CSFs are guiding the development and operational processes of the involved entities, an activity modeling language is used to describe how the CSFs *control* the involved processes in the enterprise.

The GERAM is a generalization of a number of enterprise architecture frameworks and was developed by the IFIP/IFAC Task Force on Enterprise Integration (1990-2001) (IFIP-IFAC Task Force, 2003). The purposes of the GERAM are to (Bernus, 2003a):

- *organize enterprise integration knowledge,*
- *consider all components, including human and technical aspects,*

- *apply to any complex entity (enterprises, networks of enterprises, projects, complex products etc., in any field of industry, service or government)*

The GERAM "*includes a life-cycle reference architecture (GERA), which via its modeling framework defines the scope of enterprise modeling*" (IFIP-IFAC Task Force, 2003, p.22). The GERA (Generalized Enterprise Reference Architecture) is "*the most detailed components of GERAM*" (Bernus, 2003a, p.52). It defines basic concepts in order to achieve a complete description of "*the processes in the enterprise capturing both functionality (what has to be done) and their behavior (when things are done and which sequence)*" (IFIP-IFAC Task Force 2003, pp.31-32). These fundamental concepts are (Bernus, 2003a, p.52):

- Enterprise Entity types
- Enterprise entity life-cycle
- Entity recursion
- Life history.

Figure 4 demonstrates the GERA life-cycle phases. This life-cycle can be applied for "*any enterprise or any its entities*". It includes "a *total of seven life-cycle activity types*": identification, concept, requirements, design (this can be further subdivided into two lower level types: preliminary- and detailed design), implementation, operation and decommission (IFIP-IFAC Task Force, 2003).

- The identification phase is to identify what the business is, where it is, its strategy, strategic relationships to other entities, objectives and goals of enterprise business entity (EBE) (Bernus, 2003a).
- The concept phase includes "*the definition of the entity's mission, vision, values, strategies, objectives, operational concepts, policies, business plans and so forth*" (IFIP-IFAC Task Force, 2003, p.33).
- The requirements phase is "*the activities needed to develop descriptions of operational requirements of the enterprise entity, its relevant processes and the collection of all their functional, behavioral, informational and capability needs. This description includes both service and manufacturing requirements, management and control requirements of the entity*" (IFIP-IFAC Task Force, 2003, p.33).

- The preliminary design phase is to "*specify the system design for the EBE including automated and human tasks in both the service and production, and management and control*" (Bernus 2003a, p.20).

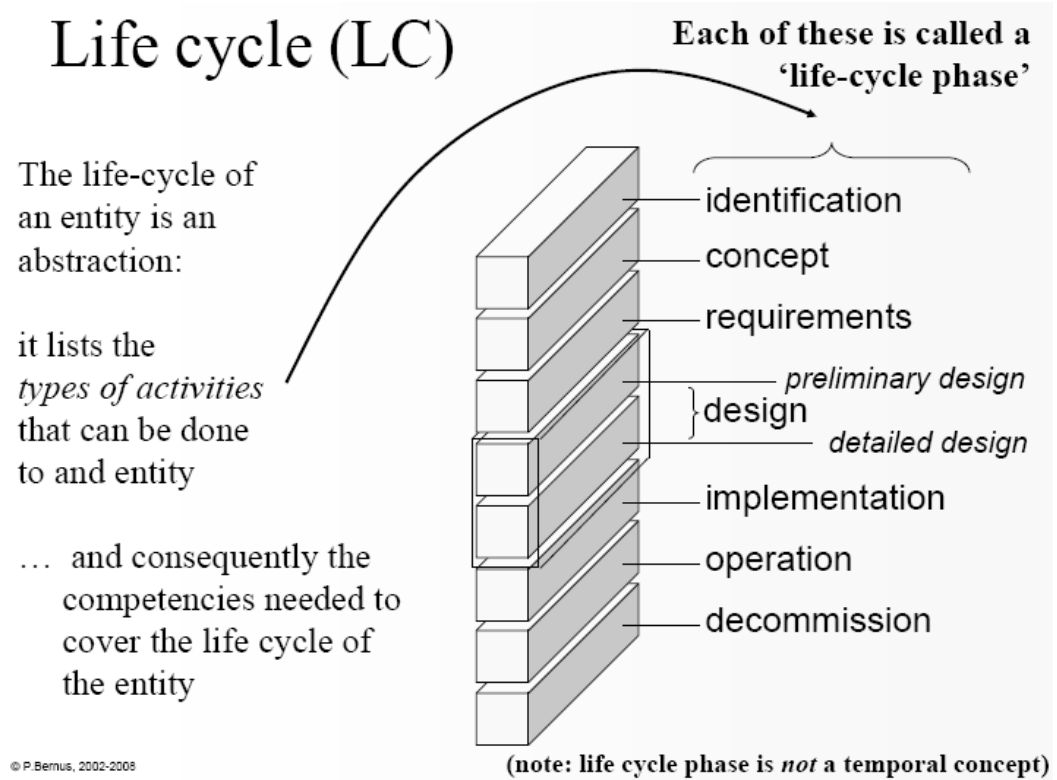

Fig. 4: GERA Life-cycle Phases
Source: Bernus, 2003a, p.54

- The detailed design phase is to design (1) "*the equipment (SW, HW) for production and service delivery*", (2) "*the human organization (tasks and job descriptions, instruction manuals, training needs, hiring guidelines, etc)*", (3) "*the management and control system (SW/HW) ERP, MIS (applications, databases, communications, etc)*" (Bernus, 2003a, p.25).

- The implementation phase includes "*the activities that define all those tasks that must be carried out to build or re-build (i.e. manifest) the entity*" (IFIP-IFAC Task Force, 2003, p.34).

- The operation phase includes "*the activities of the entity that are needed during its operation for producing the customers' product or service which is its special mission along with all those tasks needed for monitoring, controlling, and evaluating the operation. Thus the resources of the entity are managed and* controlled so as to carry out the processes necessary for the entity to fulfill its mission" (IFIP-IFAC Task Force, 2003, p.34).

- The decommission phase includes the "*activities are needed for disbanding, re-missioning, re-training, redesign, recycling, preservation, transfer, disassembly, or disposal of all or part of the entity at the end of its useful life in operation*" (IFIP-IFAC Task Force, 2003, p.34).

Figure 5 shows an example of entity recursion. Entity recursion means "*one entity may be involved in identifying, conceptualizing, specifying, designing, building, decommissioning or supporting the operation of another entity, which in tern might do the same to other entities*" (Bernus, 2003a, p.55). Additionally, Figure 6 shows an example of entity's life-history. The life-history identifies when the entity tasks need to be done (Bernus, 2003a).

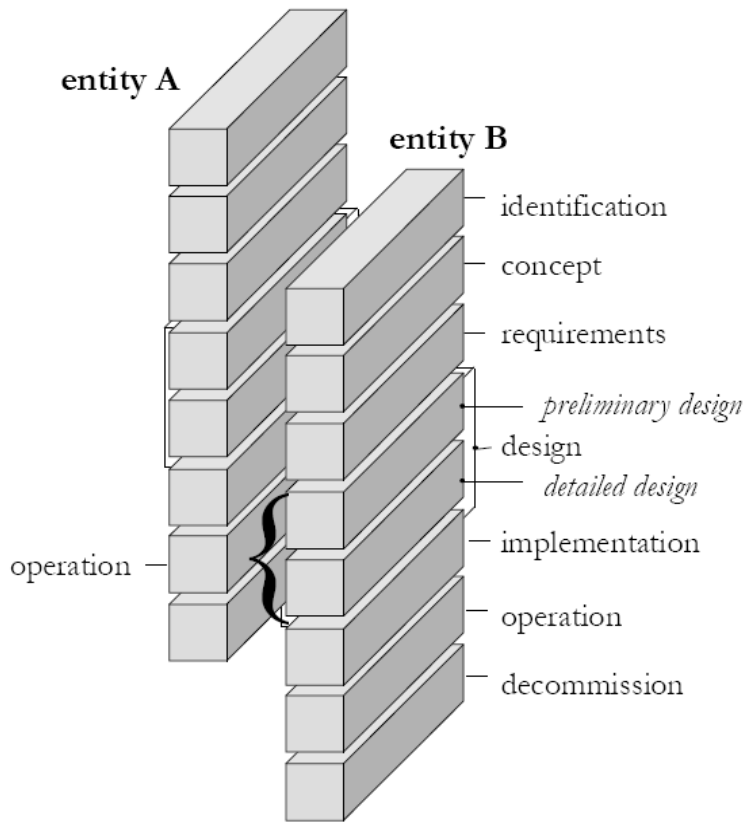

Fig. 5: Example for the relationship between life-cycles of two entities
Source: IFIP-IFAC Task Force 2003, p.36

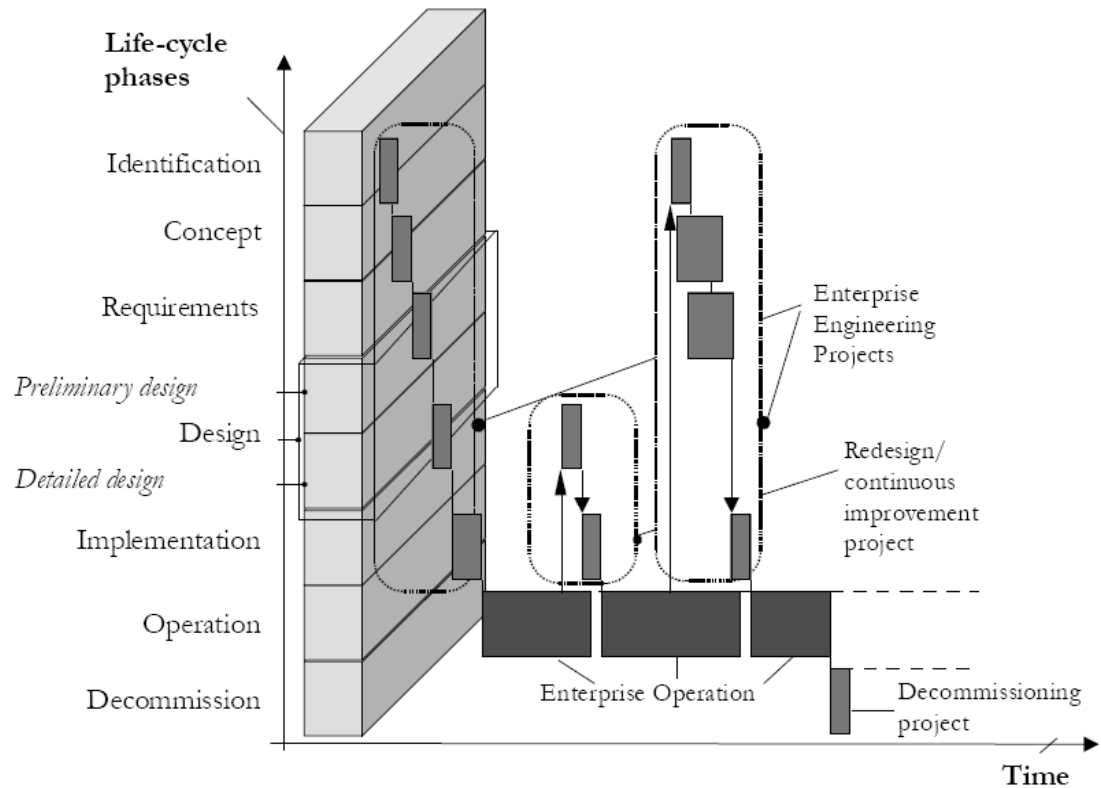

Fig. 6: Parallel processes in the entity's life-history
Source: IFIP-IFAC Task Force 2003, p.35

**GERA Modeling Framework**

GERA offers a complete analysis (which is "*based on the life-cycle content*") to define an enterprise scope through three dimensions (IFIP-IFAC Task Force, 2003, p.43):

- *"Life-Cycle Dimension: providing for the controlled modelling process of enterprise entities according to the life-cycle activities"*

- *"Genericity Dimension: providing for the controlled particularisation (instantiation) process from generic and partial to particular"*

- *"View Dimension: providing for the controlled visualisation of specific views of the enterprise entity".*

In the view dimension, GERA subdivides a model (models) into four views: content views (function, information, resource, and organization), purpose views (customer service and product, management and control), implementation view (human-, and automated tasks), and physical manifestation views (software, and hardware). In addition, other views "*may be defined according to specific user needs*" (IFIP-IFAC Task Force, 2003, p.45). Not all of these views are required to be presented in each life-cycle phase; however, "*the scope of the defined views*" is required to be covered "*by any other view subdivision*" (IFIP-IFAC Task Force, 2003). See also Fig. 7 (which demonstrates an overall diagram of the GERA modeling framework with the different views named above).

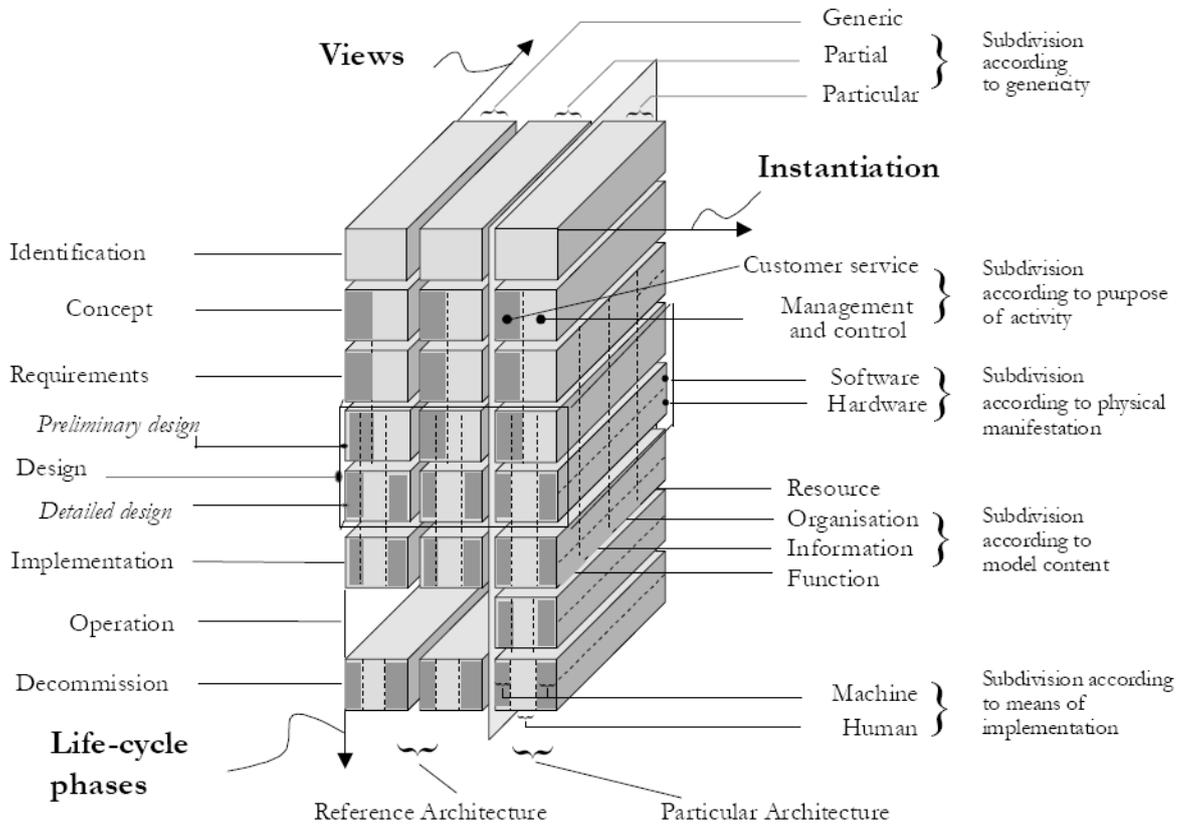

Fig. 7: GERA Modeling Framework with modeling view
Source: IFIP-IFAC Task Force 2003, p.48

For further details of the GERAM, see IFIP-IFAC Task Force (2003), which also appears as an annex to the ISO 15704:2000 standard).

The success of the B2C system will depend on both the human activities (the competencies of people who create the system and those who operate the system) and the technology (used for creating the system and the technology built into the system). The modeling framework covers all of the aspects of all of the entities involved.

Thus, the Business model in this study (Fig. 10) is based on the GERAM framework. Using the GERAM framework to develop a business model gives the following benefits:

- Gives a complete view of the B2C e-commerce organization in order to consider all involved activities in providing eService,

- Shows all involved entities in providing eService and their relationships,

- Expresses the CSFs as principles as part of the 'concept' of the involved entities; and therefore,

- Organizes and structures what needs to be done in the enterprise and its entities in order to satisfy the CSFs.

Since CSFs are guiding the development and operational processes of the involved entities, a process modelling language is used to describe how CSFs *control* the involved processes in the enterprise.

In order to describe what needs to be done to ensure that the CSFs are satisfied, an activity modelling language (IDEFØ) is used to identify the link between the processes and the CSFs. IDEFØ (pronounced 'eye-deff-zero') is a formal technique of describing processes. IDEFØ uses a box shape to define each activity. Arrows around the boxes are used to represent the information flow. The four types of arrows represent: Inputs, Outputs, Controls, and Mechanisms (or Resources). Therefore, what needs to be done (activity), who should do it (Mechanisms) and how it should be done (activity decomposition) are clearly presented and organised (NIST, 1993). See Fig. 8 below.

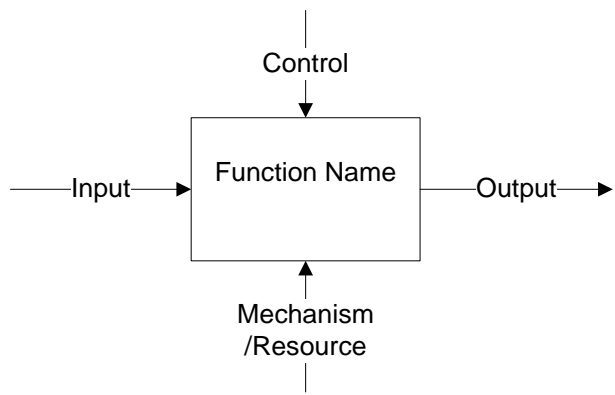

Fig. 8: IDEFØ Box and Arrows Graphics

Each IDEFØ model "*shall have a top-level context diagram, on which the subject of the model is represented by a single box with its bounding arrows*". The high-level diagram is "*decomposed into its major sub-functions by creating its child diagrams*" (NIST, 1993, pp.13-14). Furthermore, the "*functional decomposition may be performed* *through multiple levels, where sub-functions are further decomposed into even more elementary functions, etc*" (Bernus, 2003b, p.420). This structure is demonstrated in Fig. 9.

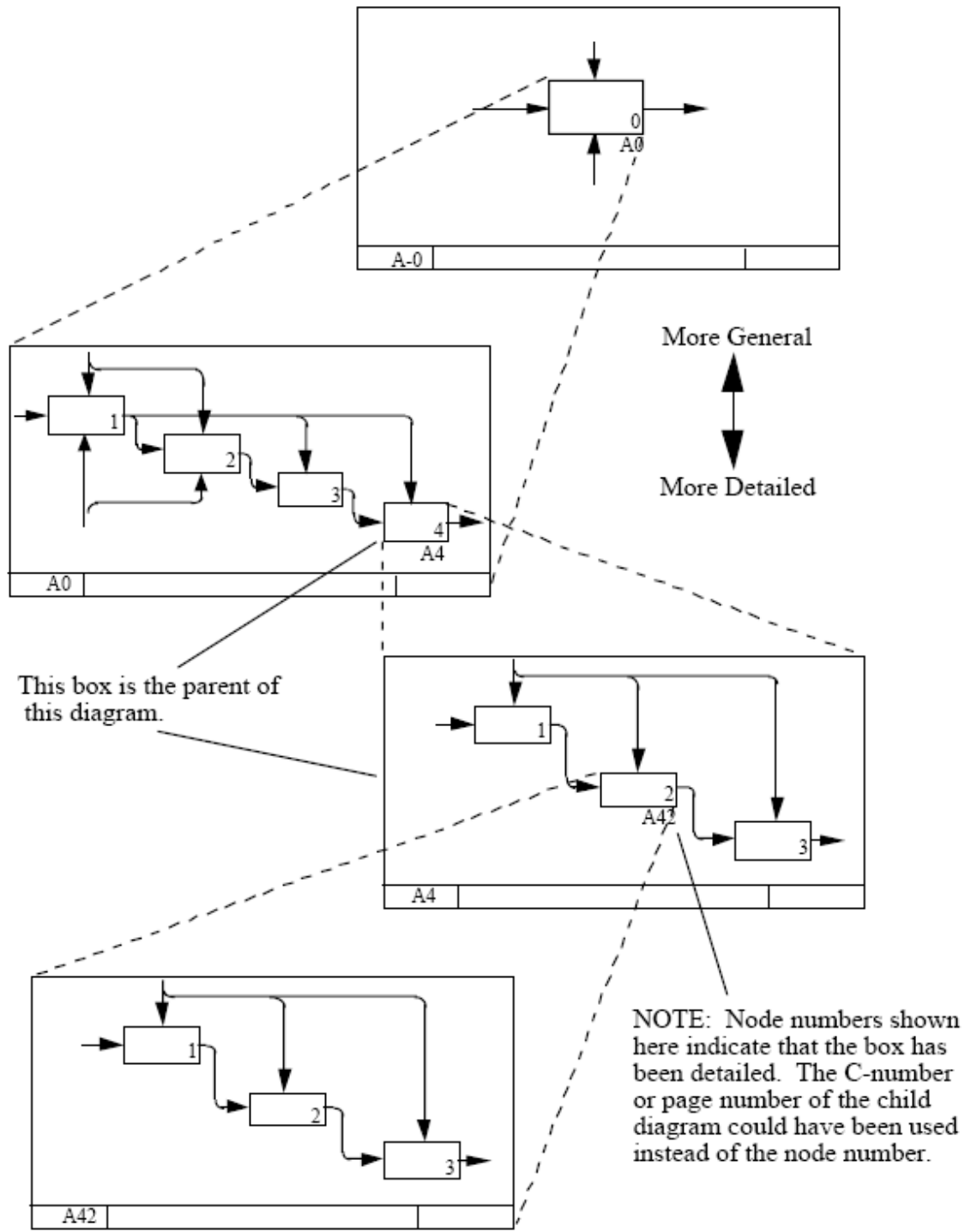

Fig. 9: IDEFØ Decomposition Structure
Source: NIST 1993, p.16

For further details of the IDEFØ modelling language, see KBSI (2006)[*].

---

[*] *IDEFØ Function Modeling Method* viewed 8 May 2008, <http://www.idef.com/idef0.html>.

# 4. Mapping CSFs onto the Business Model

In this chapter, the identified critical success factors are mapped onto the developed business model. This mapping is intended to:

(1) Demonstrate the responsibilities of the different entities and their relationships in an enterprise in order to satisfy the CSFs;

(2) Classify the CSFs in a new structure to be able to concentrate on the underlying system properties from which the rest of the properties emerge; and

(3) Help to ensure that the CSFs are satisfied by designing a process and identifying the different responsibilities in that process for the different areas of the enterprise.

## 4.1. Description of the business model

As discussed in chapter 3, the diagram of the business model (see Fig 10) describes the life cycle relationships among participating Enterprise Business Entities (EBEs). An explanation of the different responsibilities and strategic relationships among these entities is given below.

### 4.1.1. The Business Model

Figure 10 contains 12 entities: Headquarters (HQ), Business Units, Business Development, E-service Development Project (EDP), B2C eService, eService Applications, Data Service, External Services, Logistics, Manufacturer, Tangible Product and Household (Customer). The roles of these entities regarding the initiation and implementation of a B2C eService are as follows.

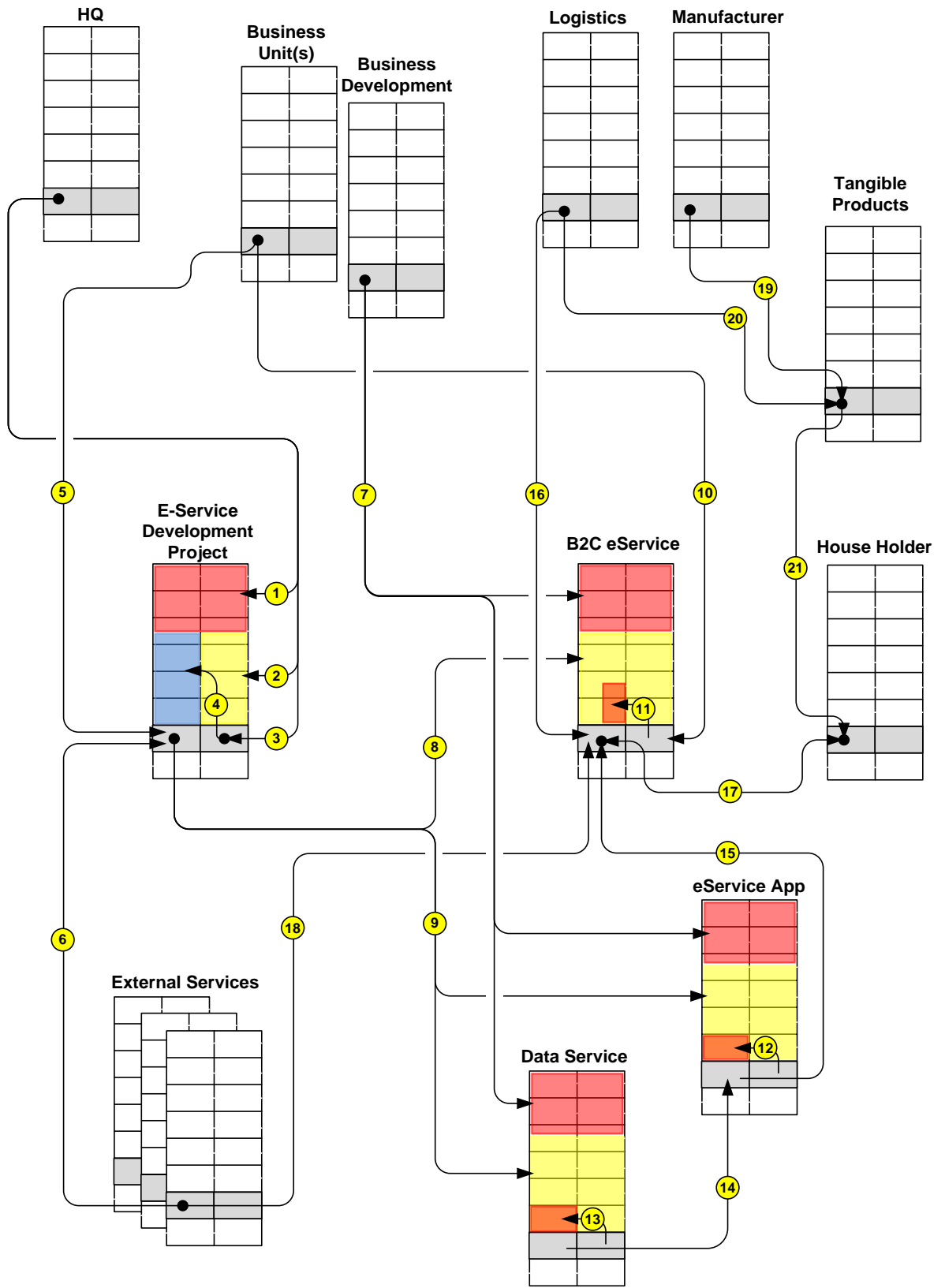

Fig. 10: Business Model

- **HQ** acts as the leader of an organization. HQ is a strategic entity that develops business objectives, such as (in this case) to develop a B2C eService for the enterprise. In this hypothetical scenario, a Business Development unit would already have developed the Concept of a potential B2C eService. The concept includes B2C eService principles, policies, values and the CSFs. Therefore, HQ can define the mandate of an E-Service Development Project, and the policies and principles that guide the execution of that project (1). It designs the management side of E-Service Development Project and allocates a project manager (2). It also provides project governance for the E-Service Development Project (3). (See Fig. 11)

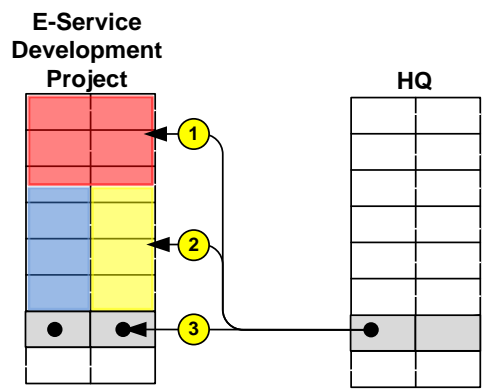

Fig. 11: HQ Relationships in the Business Model

- **Business Unit(s)** participate in the eService development project entity (5) and also provide B2C eService governance. The B2C eService processes need to be managed so that management involves traditional business decisions (marketing, purchasing, financial management etc). Business Units need to provide appropriate governance to the B2C eService and support CRM (customer relationship management) tasks (10) (see Fig. 12).

- **Business Development** is involved in the development of the B2C eService. Its responsibility is to define the Concept, and part of the Requirements (the user requirements) of three entities: B2C eService, eService applications, and data service entities (7) (see Fig. 13). The reason why the Business Development identifies only part of the requirements (user requirements) is because another part of the requirements (requirements specification) is part of the B2C eService development (which is mainly performed by the E-Service Development Project team – see Fig. 14).

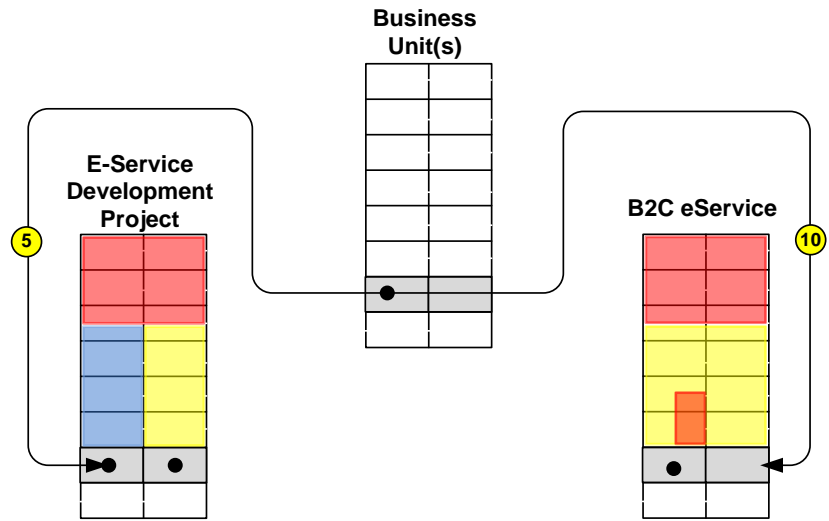

Fig. 12: Business Unit(s) Relationships in the Business Model

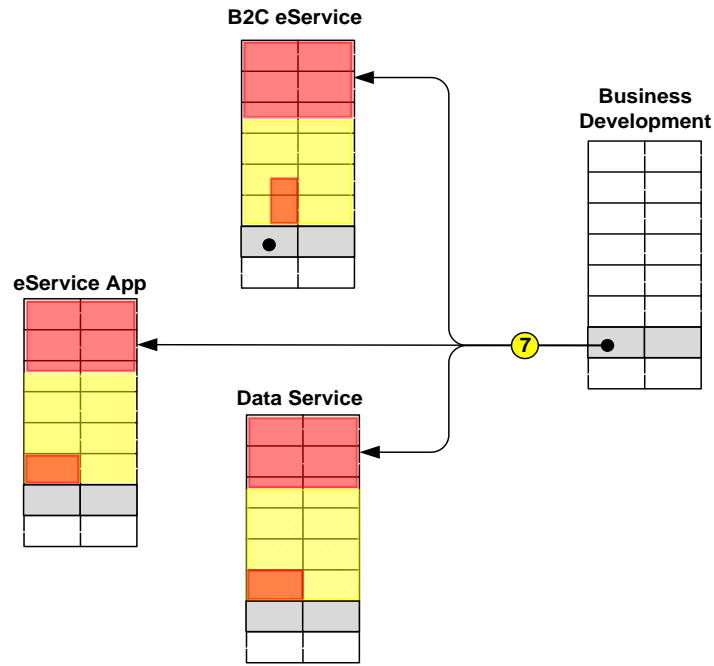

Fig. 13: Business Development Relationships in the Business Model

- **The E-service Development Project** (EDP) is responsible for the development of the B2C eService (including the development of the delivery of the eService and the development of the management of the service) and extending eService applications and data services in order to support the operation of the B2C eService. The project manager is responsible (4) for the project plan which identifies the project's processes, information, resources and any other non-functional requirements as well as having to develop the details of how the project team will operate. They also have the responsibility to manage the eService development activities, allocate internal and external project teams, and allocate external eService providers. The project team has to develop and launch the eService system (8) as well as develop the supporting application and data services (9) (see Fig. 14).

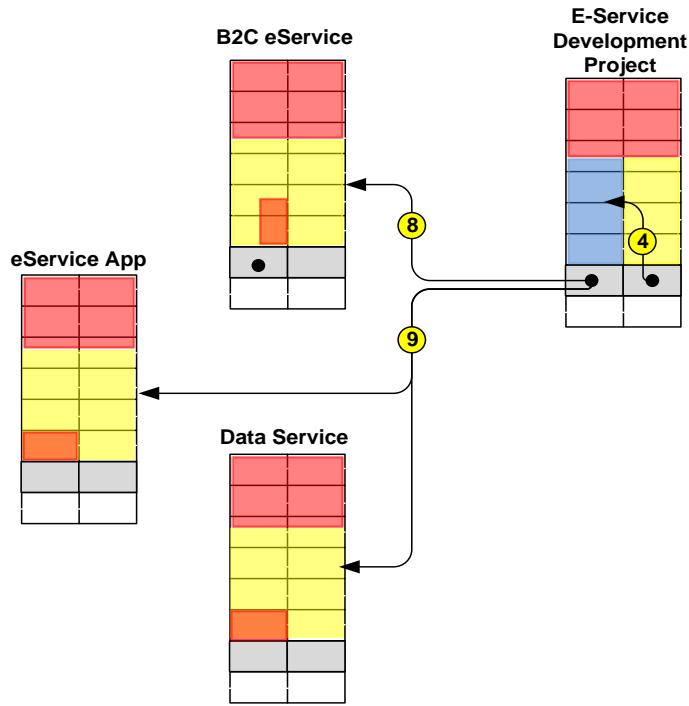

Fig. 14: E-Service Development Project
Relationships in the Business Model

- **B2C eService** represents the actual entity that interacts with online customers. Through this entity, online customers are able to do online shopping, make their enquiries, or any other transactions related to B2C services to the customers (17). The actual *provision of the B2C eService* (left-hand side of the block representing this entity) is likely to be highly (or fully) automated. In this case, the CRM tasks are supported by CRM Business Unit(s) (see Fig. 12). The B2C eService also needs to be *managed* (right-hand side of the block) which includes operational, tactical (and potentially strategic) management tasks. The management of this entity may use automated tasks, such as monitoring of performance, or performance trends, but will include humans responsible for these management tasks (11) (see Fig. 15).

- **eService Applications and Data Services** are part of the B2C eService; however, differentiating them as separate entities is due to the need to concentrate on the B2C eService, demonstrating further details of what needs to be done to deliver successful B2C eService. The role of these two entities is to support the B2C eService operation. First, the Data Service support eService Applications operation (14) and then the eService Applications support the B2C eService operation (15). What is more, the manager(s) of the eService Applications (12) and Data Service (13) has to maintain and continuously improve eService applications and data services to effectively support the B2C eService operation (see Fig. 15).

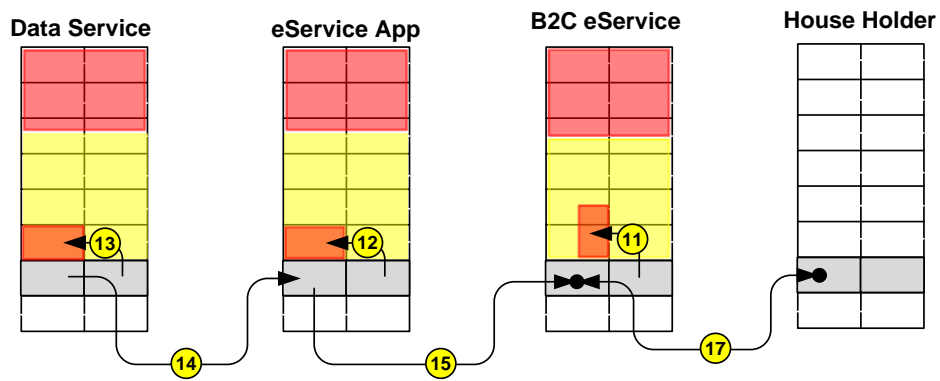

Fig. 15: B2C eService Relationships in the Business Model

- **External Services** may be involved in the development of B2C eService (6). Their responsibility is to provide technical services, such as technical consulting, software development, security, analysis and design services, and in general to participate in the eService development. Furthermore, external services would be used to support the eService (18), e.g. provide payment services for the B2C eService (see Fig. 16).

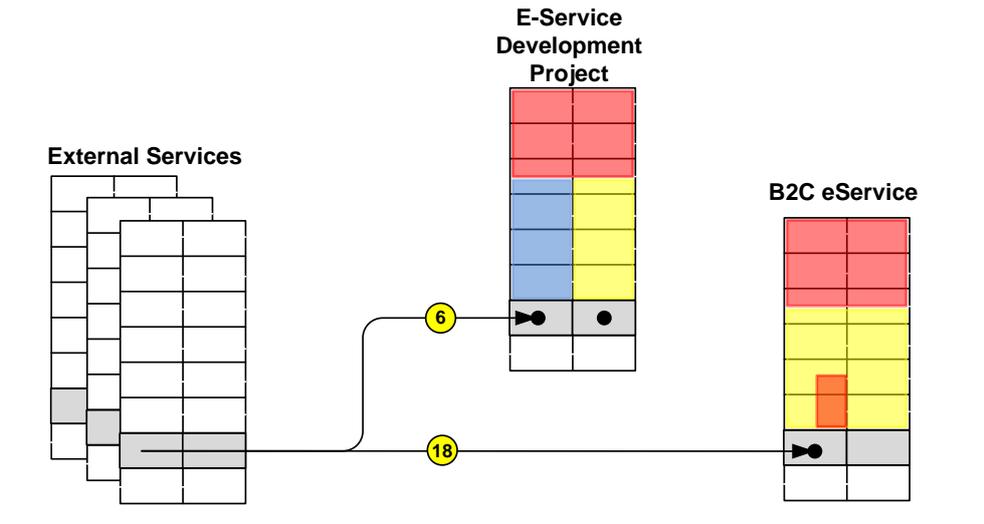

Fig. 16: External Services Relationships in the Business Model

- **Logistics** represents a logistics system interacting with shipping companies. The logistics system supports the B2C eService operation (16) to get the customers' addresses, delivery time, ordered products etc, in order to arrange with shipping companies (20) to deliver customers' orders (21) (see Fig. 17).

- **The Manufacturer** can be part of a company if they manufacture products themselves or refer to product suppliers. The role of this entity is to supply the products (19) (see Fig. 17).

- **Tangible Products.** Due to a company selling their products online, a tangible products entity is included to interact with online customers (21) through shipping companies to deliver the products (see Fig. 17).

- **The Householder** represents online customers. It might be interesting to represent online customers as a separate entity that has a life cycle. The reason why online customers are represented as a separate entity (namely 'householder') is because customers do have concepts (including policies, principles, and values), and need requirements and so on which affect the decision making to purchase online (see Fig. 17).

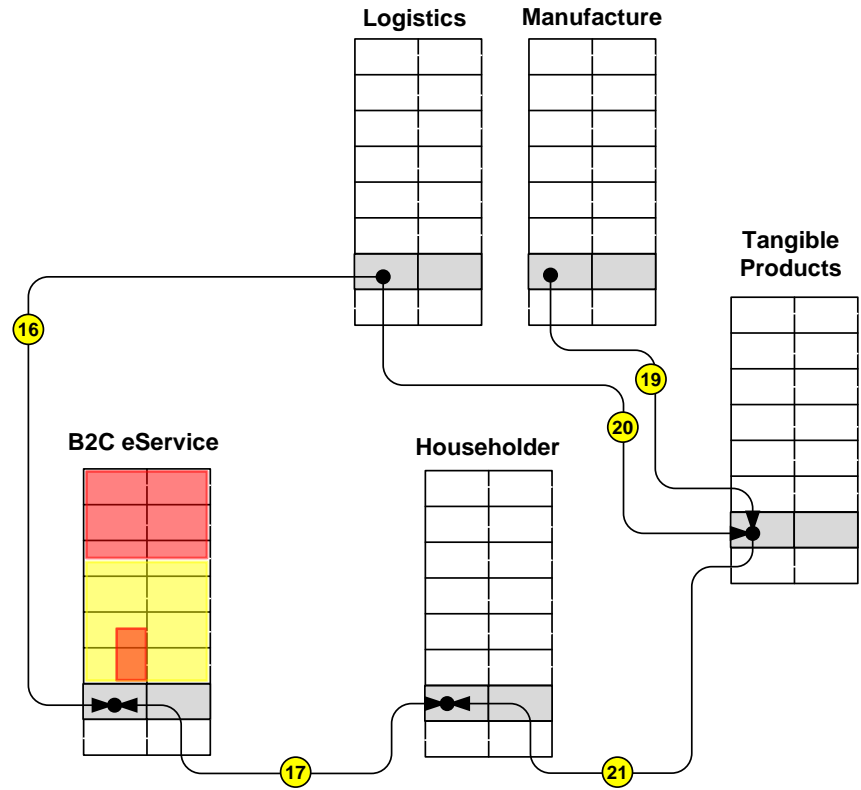

Fig. 17: Logistics, Manufacture & Tangible Product
Relationships in the Business Model

### 4.1.2. Operational and Generative Interactions

Each circled number (on the arrows in Fig, 10) refers to the explanation of a relationship among two entities. An arrow represents the fact that an entity (A) interacts with another entity (B) in one of the following ways (Bernus, 2003a):

- As (A) operates, it interacts with the operation of entity (B). These interactions are called 'operational'. This may involve the exchange of information and/or material.

- As (A) operates, it performs (or contributes to) one or more life cycle activities of (B). E.g. (A) specifies, designs, builds ... entity (B). These interactions are called 'generative'.

The explanations of these interactions are as follows:

**(1)** HQ defines the project's concept and mandate (the mandate only partially covers the requirements, i.e. it is limited to the definition of the project's tasks).

**(2)** HQ defines the requirements, architectural design, detailed design (and builds) the management side of the project.

**(3)** HQ provides project governance.

**(4)** The project manager has to define the requirements specification (processes, timing, etc), architectural design, detailed design and build phases for the project.

**(5)** The Business Unit team participates in operating the project.

**(6)** External Services providers provide technical services, such as technical consulting, software development, security, analysis and design services, to operate the project.

**(7)** The Business Development team defines the concept of B2C eService and the eService Application's and Data Service's concepts. Moreover, the Business Development defines the tasks ('user requirements') of the same entities.

**(8)** The Project team develops the requirements specification, the architectural and detailed design of the B2C eService system, implements it and releases it into operation.

**(9)** The Project team develops the requirements specification, the architectural and detailed design of the eService Application and Data Services (which support the B2C eService), implements these and releases them into operation.

**(10)** A designated Business Unit provides B2C eService governance and support CRM (customer relationship management) tasks.

**(11)** B2C eService management has to maintain and continuously improve providing eService.

**(12)** The eService Applications manager has to maintain and continuously improve eService applications to support B2C eService operation.

**(13)** The Data Service manager has to maintain and continuously improve data services to support eService applications.

**(14)** Data Service supports the eService Application's operation.

**(15)** eService Applications support the B2C eService's operation.

**(16)** There is a two-way relationship between the B2C eService and Logistics entities in order to exchange delivery information.

**(17)** Interaction between B2C eService system and its customers.

**(18)** Provide payment service.

**(19)** Manufacture and package tangible products to be ready for shipping.

**(20)** Deliver customer orders.

**(21)** The relationship between customers and products (the product supports the customer's 'operation').

## 4.2. Systematizing Critical Success Factors

In this section, the categories in Fig. 2, "Relationship Chart of the Identified B2C E-commerce Critical Success Factors", can be further structured so as to be able to concentrate on the underlying system properties from which the rest of the properties emerge.

Fig. 19 demonstrates a new structure of the CSFs, illustrating the hierarchal structure of CSFs. This structure clarifies the underlining critical success components. B2C e-commerce success is influenced by two main factors: customer satisfaction and loyalty, and enabling enterprise processes. Furthermore, customer satisfaction and loyalty, and enterprise processes are influenced by a number of factors, as discussed below.

### 4.2.1. B2C E-commerce Success through Two Dimensions

#### 4.2.1.1. Customer satisfaction and loyalty (see Fig. 19)

Customer satisfaction and loyalty are affected by three main factors (Ribbink et al., 2004): service quality, e-commerce system quality, and trust. Each one of the previous factors contains several components.

**(1)** Service quality contains five elements: payment processes, product quality, variety of goods/services, low price goods/services, and support & services. Support and services contains two factors: responsiveness/interaction, and delivery of goods/services.

**(2)** E-commerce system quality comprises three factors: speed of system, usefulness, and website design. Website design has three factors: content quality, customization, and ease of use.

**(3)** Trust is a result of system security, information privacy, social presence and ease of use.

### 4.2.1.2. Enterprise processes (see Fig. 19)

Enterprise processes involve what an e-commerce organization does to ensure that the quality of the internal processes is satisfied. Internal processes include nine factors, which are: Electronic Commerce Customer Relationship Management (ECCRM), e-commerce strategies, evaluation of e-commerce operation, technical e-commerce expertise, low cost operation, end-user participation in development, combining e-business and value proposition, monitoring internal processes and competitive activities, and replication of offline brand.

Note that organizing these factors, as seen in Fig. 19, hides the fact that these factors are not independent. For example, trust *might* be influenced by the e-commerce system and service quality (although the literature does not identify this as an important relationship). Organizing factors in this way gives a clear and easy understanding of the *main* relationships among these factors.

### 4.2.2. Definitions (explanations) of identified CSFs

Explanations of the identified CSFs are defined in terms of B2C e-commerce system as follows.

- **Content Quality** "*refers to the characteristics and presentation of information in the e-commerce system*" (Molla & Licker, 2001, p.137).

- (Continuous) **Evaluation of E-Commerce Operation** supposes the measurement and monitoring of Performance and Value Delivery (Sung, 2006; Dubelaar, Sohal & Savic, 2005).

- **Customer Loyalty** is "*a deeply held intention to repurchase a preferred product/service consistently from a particular e-vendor in the future, despite the presence of factors or circumstances that may induce switching behavior*" (Yeo & Chiam, 2006, p.334).

- **Customer Satisfaction** is "*the perception of pleasurable fulfillment of a service*" (Shankar, Smith & Rangaswamy, 2003, p.154).

- **Customization** refers to the website personalization that "*addresses the ability to offer content tailored to the preferences of each user*" (Lepouras & Vassilakis, 2006, p.1).

- **Ease of Use** refers to customers requiring "*less cognitive effort*" and having more attention to engage with the system, "*particularly with a friendly and personable online agent*" (Chen et al., 2007, p.76).

- **E-commerce (EC) Strategies** include strategic management tools, marketing, sales, after sales services etc. (Chen et al., 2006).

- **E-Commerce Customer Relationship Management (ECCRM)** is "*the activities related to initiating, negotiating, and executing business transactions online*" (Schodeler & Madeja, 2004, p.38).

- **E-commerce System Quality** includes various dimensions: system reliability, accuracy and flexibility, time of response, ease of use, 24 hour availability, page loading speed, stability of the software and hardware, architecture of the system, visual appearance and accessibility (Molla & Licker, 2001).

- **Goods/Services Delivery** concerns the accuracy of ordered product and promised delivery time (Sung, 2006).

- **Information Privacy** "*relates to the concern about the potential misuse of personal information by marketers*" (Ribbink et al., 2004, p.448).

- **Low Cost Operation** of the actual system (Sung, 2006).

- **Low Price of Goods and Services** provided to customers (Sung, 2006).

- **Monitoring Internal Processes and Competitor Activities** (Dubelaar, Sohal & Savic, 2005).

- **Payment Processes** includes payment security and having various methods available for customers to pay (Sung, 2006).

- **Replication of Offline Brand** means selling products that can be found or known offline (Dubelaar, Sohal & Savic, 2005).

- **Responsiveness/Interactions** from both the system-based assistance and the human-based assistance in order to complete customers' tasks (Aberg & Shahmehri, 2000).

- **Social Presence** is the "*extensive ongoing interactions that enable individuals to create reliable expectations of what other persons or organizations may do*" (Gefen & Straub, 2004, p.410).

- **Support and Services** "*can take different forms and may include the following: site intelligence (the extent to which the e-commerce system remembers repeat users and aids them in achieving goals), relevant search facilities, feedback, calculators, currency converters, tracking order/shipment status, account maintenance, payment alternatives, FAQs, etc*" (Molla & Licker, 2001, p.138).

- **System Security** "*concerns the risk of third parties obtaining critical information about the customer*" such as "*access to credit card or bank account details*" (Ribbink et al., 2004, p.448).

- **Technical E-commerce Expertise** on both sides: e-commerce experts and necessary e-commerce technologies (Sung, 2006).

- **Trust** is "*a social complexity-reducing mechanism that leads to a willingness to depend on a vendor; this willingness is derived from the perception that the vendor will fulfill its commitments*" (Gefen & Straub, 2004, p.410).

- **Usefulness** is defined in general as "*the degree to which a person believes that using a particular system would enhance his or her ... performance*" (Chen & Yang, 2006, p.290).

- **User Participation** and involvement in systems development activities (Terry & Standing, 2004).

- **Website Design** includes various aspects (colours, font size, graphics, response time, pages layout etc) that play an important role determining its usability (Green & Pearson, 2006).

### 4.2.3. Three Aspects to each CSF

Each of the identified CSFs has three aspects to it (see Fig. 18):

1. The quality of B2C eService which refers to a CSF itself (i.e. the quality that the B2C system must have).

2. The quality of the eService Development Project that designs and implements the service (which service is mainly delivered by an automated system).

3. The quality of the management (strategic, tactical and operational management, which includes supervision/monitoring, maintenance and continuous improvement) of the B2C Service (these are mainly a human tasks, but supported by technology).

Notice that the last two aspects (the quality of eService development project and the quality of the B2C service's management processes) do not have to place equal weight on all CSFs. For example, the eService development project (which designs and implements the service) needs more focus on the quality of the payment processes than on product quality, whereas maintaining good product quality relates to the quality of the B2C service management processes (in order to ensure the quality of product offered at any time).

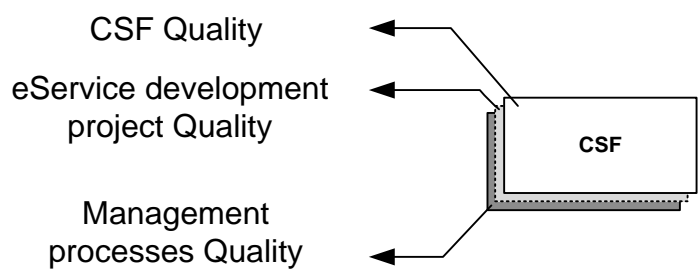

Fig. 18: CSFs Quality Aspects

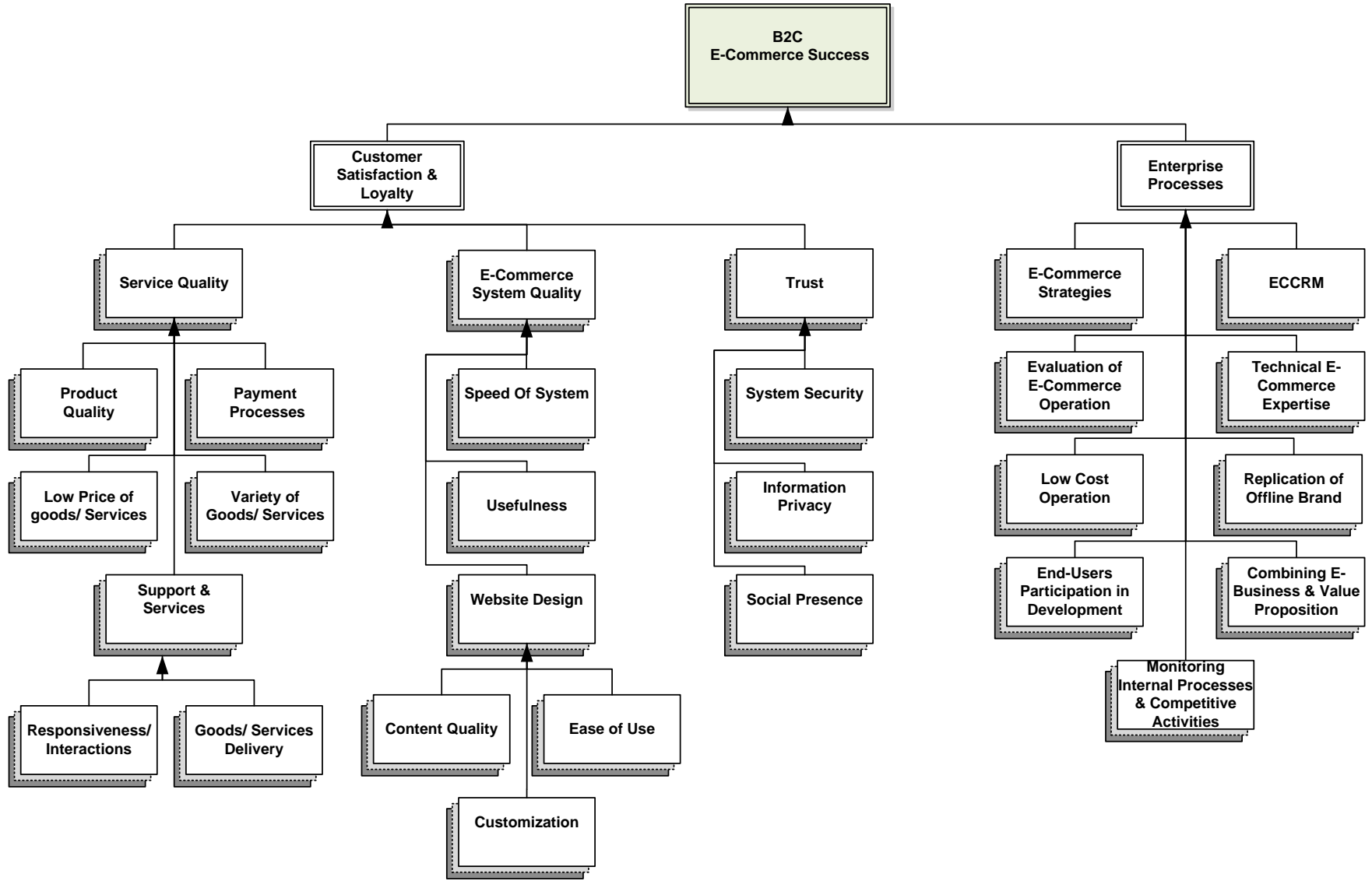

Fig. 19: The New Structure of the Identified B2C E-commerce System CSFs

### 4.3. Satisfying CSFs: Ease of Use (as a case)

This section discusses what needs to be done to ensure that the CSFs are satisfied. In order to structure and organise these tasks, a process model of B2C Service Development has been developed, concentrating on the necessary activities, their outcomes and input/output relationships, as well as identifying the role of Enterprise Business Entities (EBEs) in this process (for the list of the typical EBEs, see Fig. 10).

Through a better understanding of the B2C service development process, the CSFs of the eService development project as well as its relationships to the CSFs of the B2C Service will be clarified.

In order to understand the above process, an activity modelling language (IDEFØ) is used. IDEFØ is a language that was developed for exactly this purpose (NIST, 1993), as may be understood from looking at IDEFØ Box and Arrows Graphics (see Fig. 8 in section 3.2).

Before starting to analyze (using IDEFØ) what activities need to be performed in the development process and what information (such as standards) must be taken into account to satisfy the CSFs, one must start with the identification of each CSF, and determine to which entity it belongs as a responsibility of that entity (see Fig.10). For example, a CSF of the B2C system must be ensured through a series of activities in the life cycle of the B2C system, where these activities are performed by the B2C development project (i.e. the process of the project). What needs to be done in this project is discussed first using a high level IDEFØ diagram. Then, the diagram is decomposed demonstrating further details. In the decomposition of each life cycle activity, a brief discussion of the role of the CSFs will be included, and the chosen CSF and the necessary activities to satisfy it will be discussed.

Due to the limited scope of this thesis, one example CSF is chosen and discussed, namely 'ease of use' (or in general 'usability'), as a critical success factor of the B2C e-commerce system. All other CSFs could be analyzed in a similar manner, and the developed life cycle process model of the involved entities could be extended with the necessary activities (and their information inputs or controls) to satisfy these CSFs.

In terms of electronic customer relationship management, ease of use refers to customers requiring "*less cognitive effort*" and having more attention to engage with the system, "*particularly with a friendly and personable online agent*" (Chen et al., 2007, p.76). For customers' online interaction with B2C e-service, ease of use is a key component. However, this needs to be translated into system characteristics of the B2C eService entity. Thus, ensuring that this translation happens, the following questions need to be answered: 1) what system characteristics are needed in the B2C e-commerce system? 2) What do developers need to do, so that the developed system will have these characteristics? It is necessary to analyze what factors contribute to the ease of use and how such factors are satisfied through *appropriate life cycle activities* of the involved entities. The EDP is the central entity which takes the main role in developing a B2C eService system. HQ is involved in initiating the B2C eService development project and developing the project mandate (see Fig.10), and the Business Development Unit develops the B2C eService design brief. Given that the main interest in this thesis is to understand the processes involved in the EDP, the EDP processes are separated into the following two aspects: a) manage eService development and b) develop the eService system.

### 4.3.1. Managing the eService Development

This part involves defining a plan for the project and allocating internal and external project teams, service providers, and technical resources.

Figure 20 demonstrates a high level view of this management process. This figure gives an overview of management's role in initiating and delivering a B2C eService system. The process is called 'Manage eService Development' and referenced 'A-Ø' (the usual code for an IDEFØ 'context diagram'). HQ and the Business Development team need to manage the eService development by taking into account the business intention (e.g. eService strategic objectives). From the point of view of usability, the management of eService development must take into account a number of relevant inputs, such as standards: ISO 9126 (Software Quality-see Fig. 23), ISO 9241-11 (Guidance on Usability), ISO 13407 (Human Centered Design Process for Interactive Systems), potential B2C solutions (existing products that can be used to build a system), and general Human Computer Interaction (HCI) design guidelines. The outcomes of this management activity include a preliminary and a final implementation plan, and allocated internal and external project teams and external service providers.

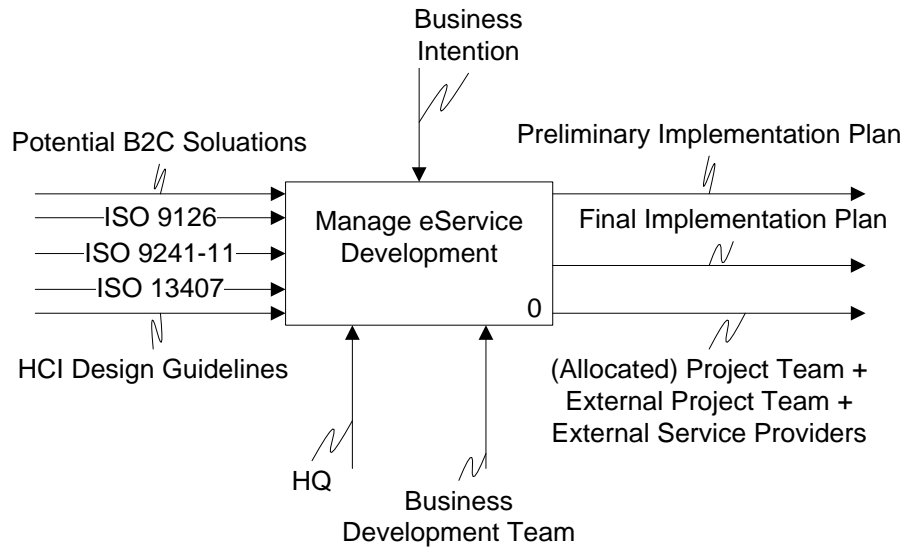

Fig. 20 (A-Ø) Manage eService Development

Figure 20 is further deconstructed demonstrating details of activities involved in managing eService development (see Fig 21). The process starts with the definition of the B2C 'concept' which (according to the used enterprise architecture framework) includes principles, policies, values, tasks, user requirements, and CSFs. This 'concept' is summarized in a 'B2C eService design brief' document. Given a design brief and a business intention, HQ is able to initiate the B2C eService development project. This initiation is to allocate a project manager and to provide a project mandate (tasks) and project principles, policies, and a project budget. Given the project mandate and the B2C eService design brief, the allocated project manager and HQ are able to develop a project plan. Since usability is the focus here, relevant usability inputs are needed in the development of a project plan. Developing a project plan requires taking ISO 13407 (Human Centered Design Process for Interactive Systems) into consideration. This project plan needs to include the definition of the project's processes, information, resources and requirements. Given all of these, HQ, project manager and Business Development team are able to jointly manage the eService development project. The outcomes of managing the eService development project are the allocation of internal and external project teams and external service providers, and a preliminary implementation plan. Furthermore, the final implementation plan is also an outcome, but to be able to deliver this plan, it is necessary to have a proposed B2C master plan (which comes from the developers – see A5 in Fig. 21). The proposed B2C master plan is the outcome of the architectural design phase of developing the B2C eservice system (see section 4.3.2). Finally, given the final implementation plan, the Business Unit team and allocated internal and external teams will develop a B2C eService system and release the eService system to operation. Respective parts of the B2C eService system development are controlled by the B2C eService design brief, preliminary implementation plan, and the final implementation plan. The B2C eService system development is a separate process and its details are further discussed in section 4.3.2. The only reason why A5 appears in Fig. 21 is to show the relationship between the management process and the development process.

Notice that all necessary activities that ISO 13407 requires will be taken into account in the final implementation plan (indirectly influencing the development process). This is necessary, because the standard provides advice on meeting "*quality in use by incorporating user centered design activities throughout the life cycle of interactive computer-based systems*" (Bevan, 1999, p.2). This standard defines "*user centered design as a multi-disciplinary activity, which incorporates human factors and ergonomics knowledge and techniques with the objective of enhancing effectiveness and productivity, improving human working conditions, and counteracting the possible adverse effects of use on human health, safety and performance*" (Bevan, 1999, p.2).

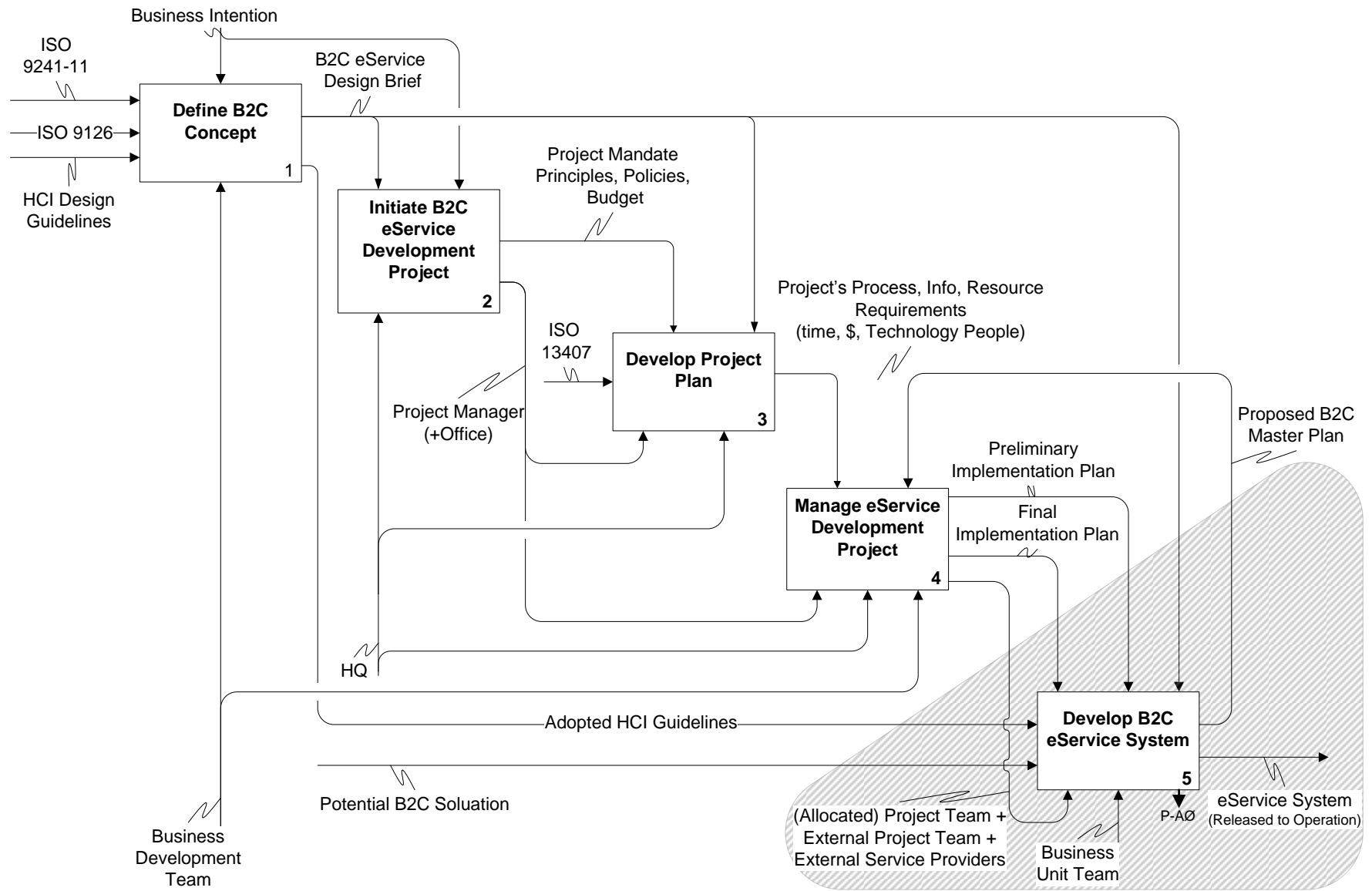

Fig. 21: (M-AØ) Manage eService Development

The definition of the B2C eService concept is further deconstructed (see Fig. 21 and Fig. 22). Figure 22 shows the breakdown of defining the B2C 'concept'. According to enterprise architecture frameworks, the definition of the 'concept' involves three important activities: defining design principles, policies and tasks. Given the business intention, and the listed standards (the relevant standards for usability), the Business Development team is able to define the B2C eService principles and policies. The outcome of the B2C eService principles definition is to provide design principles, values and CSFs; and the outcome of the B2C eService policies definition is the design policies. These outcomes are joined together to give the B2C eService 'concept' (which will become part of the design brief). Given the B2C eService 'concept' and business intention, the Business Development team is able to define the B2C eService tasks to provide tasks of B2C eService and its management. Finally, joining the B2C eService tasks and 'concept' gives the B2C eService design brief.

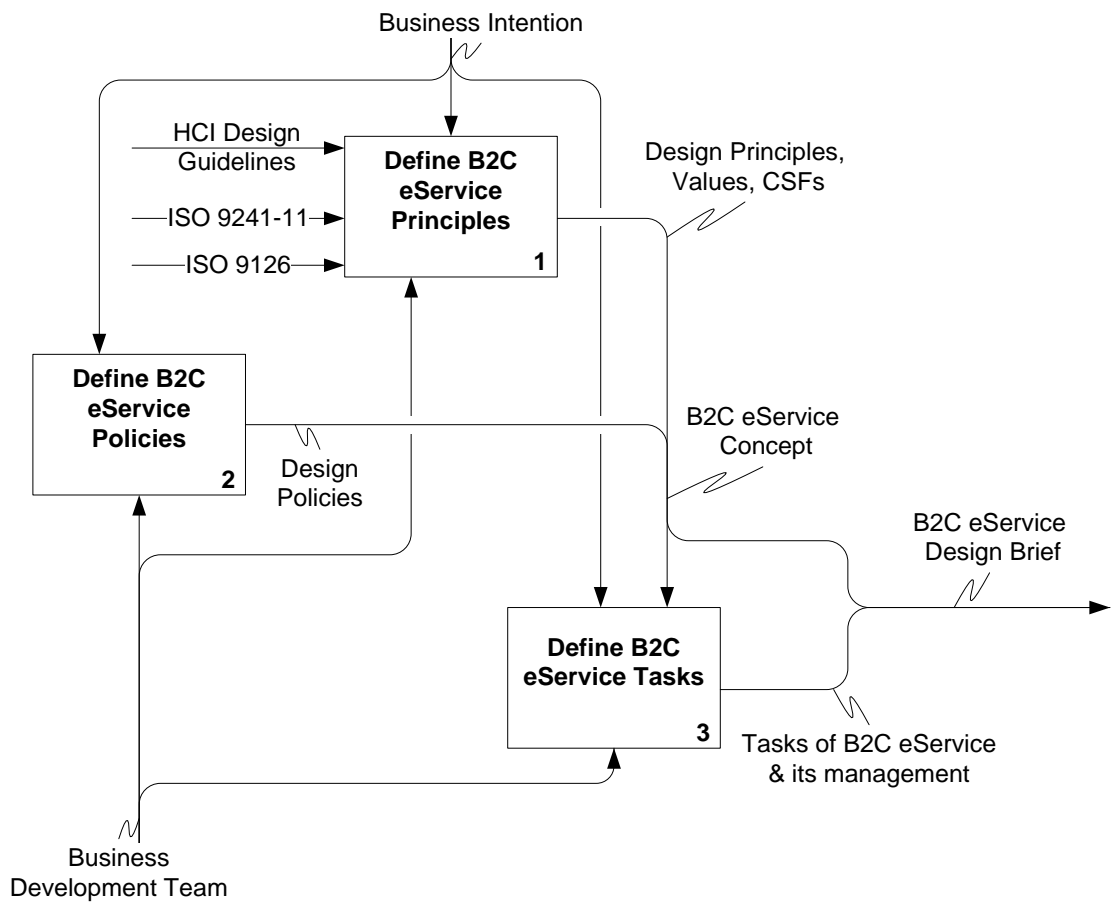

Fig. 22: Define B2C eService Concept

As Fig. 22 shows, the definition of the B2C eService principles takes various standards as inputs. ISO 9126 (Software Quality – see Fig. 23) and ISO 9241 part 11 (Guidance on Usability) can be used to adopt principles for usability. ISO 9126 is an international standard for evaluating software quality and provides quality characteristics and guidelines for usability (Stefani & Xenos, 2007). ISO 9241-11 is a part of the international standard ISO 9241 series which also provides guidance on usability (ISO, 1998). What is more, the definition of B2C eService policies does not have defined inputs (in this presentation), because management needs to decide this detail based on the intention of the business (e.g. mandate the use of commercial off the shelf products, or allow internal system development). In addition, government legislations and rules, and customers' cultures are all important issues which need to be taken into account and reflected in these policies.

ISO 9126 defines software quality characteristics to be used when specifying system requirements and to be enforced throughout the life cycle of the B2C system. The standard defines six quality characteristics:

- Functionality
- Usability
- Efficiency
- Reliability
- Maintainability
- Portability

Each characteristic property has quality sub-characteristics (see Fig. 23) (Stefani & Xenos, 2007); details of these characteristics will be explained in section 4.3.2.1.

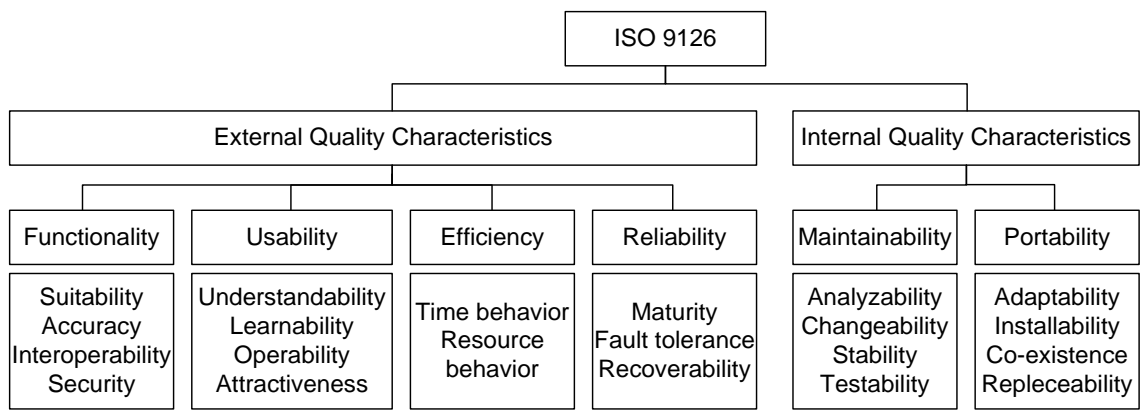

Fig. 23: Software Quality – ISO 9126
Source: Abran et al., 2003, p.328

ISO 9241 part 11 contains guidelines for achieving the quality of usability and identifies necessary information to be taken into consideration when "*specifying or evaluating usability of a visual display terminal in terms of measures of user performance and satisfaction*" (ISO 9241-11, 1998, p.5). The standard specifies and measures usability through its components. Usability is broken down into four components:

- Effectiveness
- Efficiency
- Satisfaction
- Context of use

'Context of use' can be further divided into four sub-components: user, task, equipment and environment (ISO 9241-11, 1998). Further details of using this standard will be explained in section 4.3.2.1.

### 4.3.2 Developing an eService System

This activity is the actual development of the B2C eService system. It involves the participation of two entities: 1) Business Development Unit to confirm the B2C eService design brief and 2) E-service Development Project (EDP) entity to design and develop the system. Figure 24 demonstrates a high level diagram of the eService system development. Given the B2C eService design brief, preliminary and final implementation plans, ISO 9126, and ISO 9241-11, internal and external teams are able to jointly: 1) propose a B2C master plan (in order to get a confirmed implementation plan from management), and 2) deliver the B2C eService system. Moreover, it is necessary to adopt the HCI guidelines and find potential B2C system solutions in order to deliver the B2C eService system. The focus on which of these two inputs is more important depends on which approach the company wants to adopt (as determined in the design policies). Thus, the focus on adopting HCI guidelines means either developing a totally new solution (which requires a professional team to develop it), or in contrast, focus on adopting potential B2C system solutions (using existing e-commerce tools provided by other companies, such as eBay). Regarding these alternatives, Roben (1998) suggest that an e-commerce company needs to emphasize the main purpose of their website. That is to say, "*is the site meant to be self-sufficient, or is it meant to reduce demands on the staff? Is the site only an e-Commerce site, or is it part of a larger company site that conveys more information*" (Roben, 1998, p.112). The adoption of a ready solution is suggested, where if the e-Commerce is a part of a company comprehensive, an internal development is more appropriate. The choice of these two alternatives is a part of the policy development.

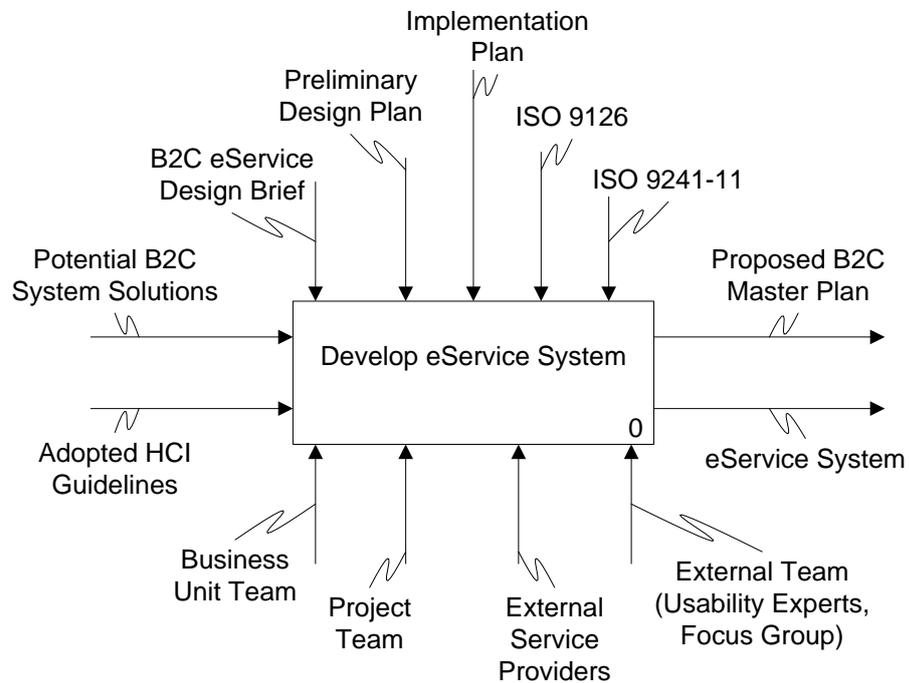

Fig. 24: (A-Ø) Develop an eService System

Figure 24 is further divided into two main processes: initiate and implement the eService system (see Fig. 25).

**(1)** Initiating the eService system involves the development of a Master Plan that is suitable for deciding the feasibility and cost of building the B2C system. This activity is controlled by the B2C eService design brief, preliminary implementation plan, and after receiving feedback from management, the final implementation plan. Further feedback may come from the implementation (A2 in Fig. 25), 'implementation feedback & change requests' (e.g. due to the results of the usability tests). This initiation requires information about potential B2C system solutions and the adopted HCI guidelines. The outcome of this process is: 1) a proposed B2C system master plan (to be reviewed by management), and (in the end) 2) an approved B2C system master plan (to be used for implementation).

**(2)** Implementing the eService system requires an approved B2C system master plan, as well as having to take into account standards as noted (ISO 9126 and ISO 9241-11) to guide the project participants (Project Team, the Business Unit team, and usability experts) whose common task is to deliver the eService system.

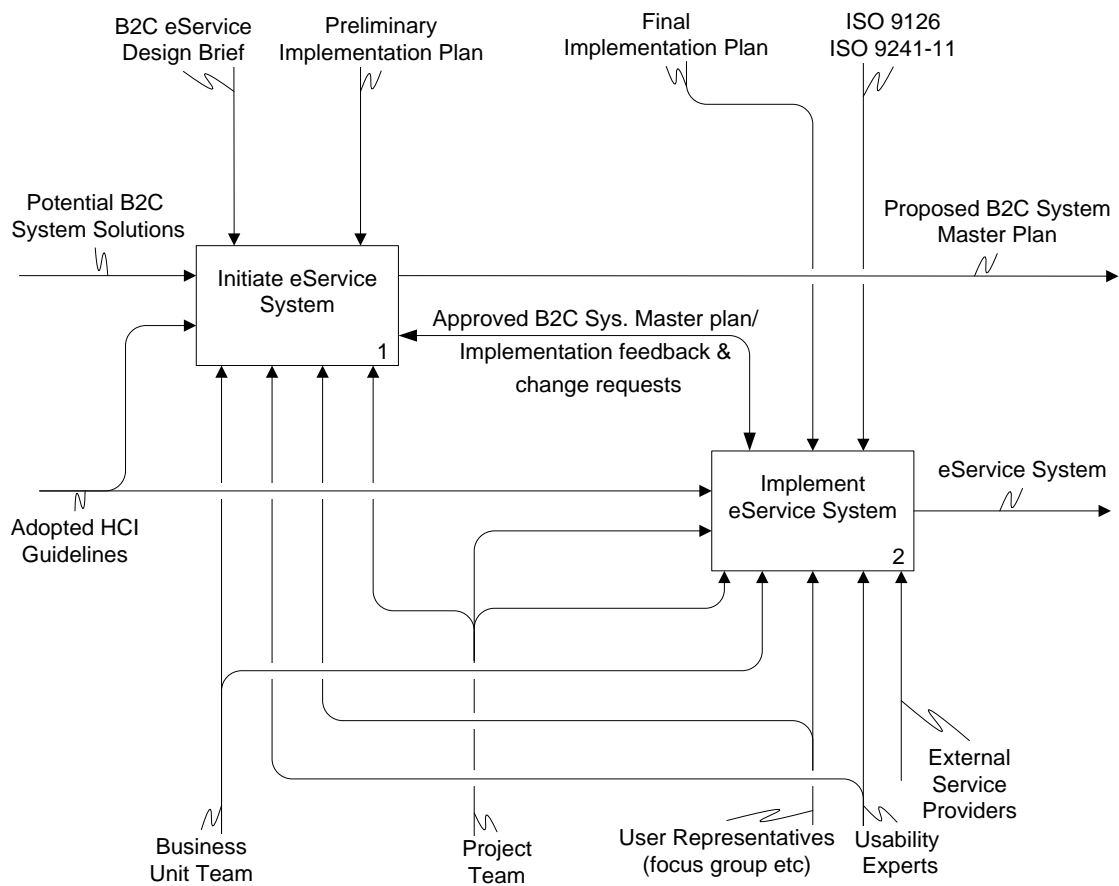

Fig 25: (P-AØ) Develop an eService System

### 4.3.2.1 Initiating the eService System

Initiating the eService system includes the following three sub-activities (see Fig. 26): confirming the design brief, defining the B2C requirements and architectural design. The process starts with confirming the design brief that was provided by the management level (see Fig. 21). Confirming the design brief is guided by the B2C eService design brief and preliminary implementation plan, and is a responsibility of the EDP (see Fig. 14). This activity is important, because the project participants must ensure that they have understood correctly what management wanted. Given the confirmed design brief, the preliminary implementation plan, and relevant standards (ISO 9126 and ISO 9241-11), the project team is able to define B2C requirements (with the participation of user representatives or experts, as needed). The B2C eService requirements (which document also includes the confirmed design brief) and the preliminary implementation plan guide the architectural design. The architectural design is to define the aggregation of hardware and software elements that make up the complete B2C system. Architectural design is performed by the project team. The project team needs to take HCI architectural design guidelines and potential B2C system solutions into consideration to propose a B2C system master plan. The proposed B2C system master plan needs to be reviewed by HQ, the project manager, and the Business Development team, as discussed in section 4.3.1 (see Fig. 20). This revision enables the project team to produce an approved B2C system master plan. All these three phases (confirm design brief, define B2C requirements, and define architectural design) are under the control of required changes that may be proposed by usability tests. A usability test is performed in the implementation of the eService system (see section 4.3.2.2 and Fig. 32). In the next section, the details of requirements definition and architectural design are discussed.

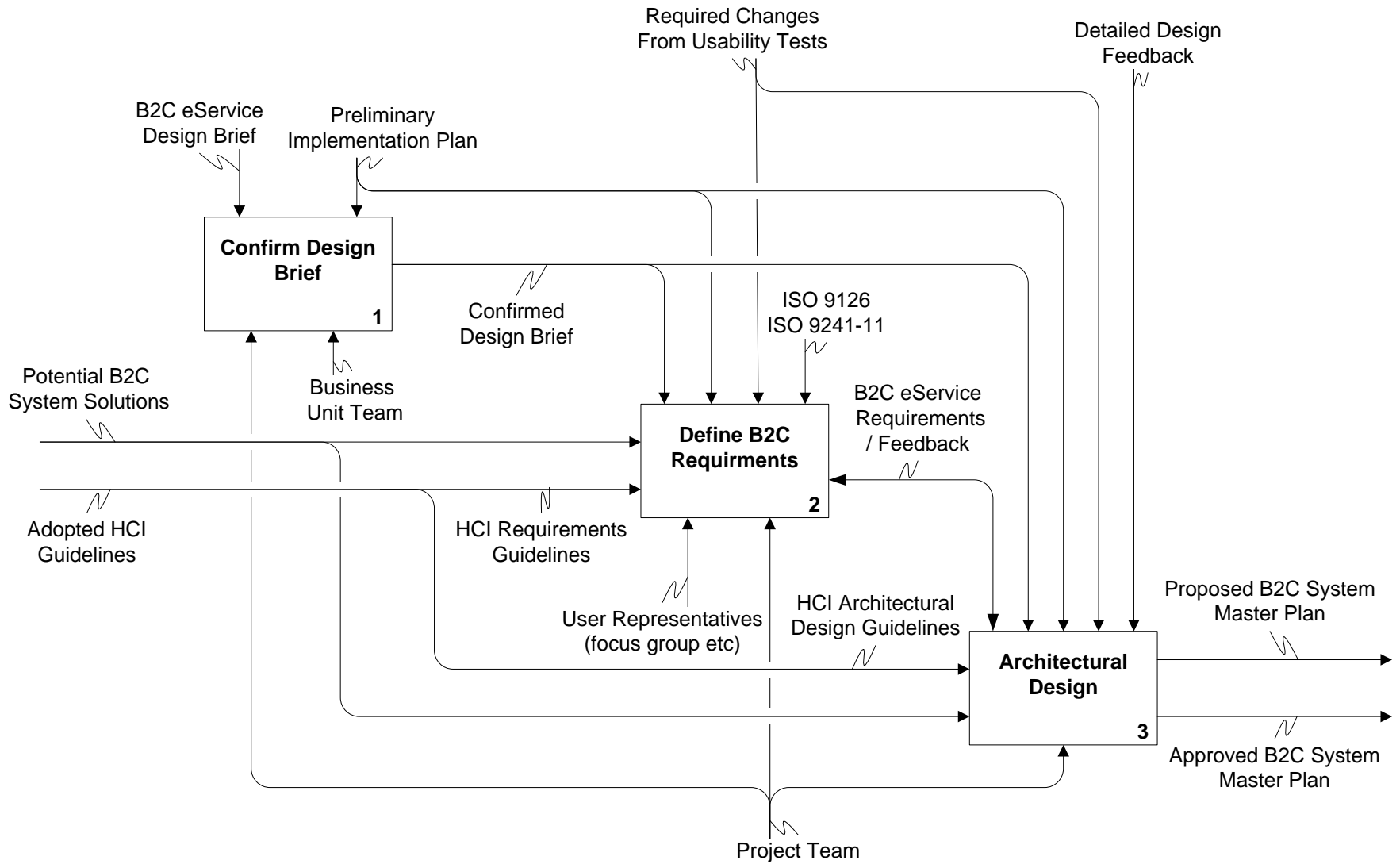

Fig. 26: (P-A1) Initiate the eService System

**Define B2C Requirements**

As Fig. 19 (in section 4.2.3) shows, the most critical success factors, including ease of use, relate to the e-commerce system and service qualities. For this reason, the requirements definition is based on the ISO 9126 quality model. ISO 9126 defines the software quality as "*a set of features and characteristics of a product or service that bear on its ability to satisfy stated or implied needs*" (ISO/IEC 2001, cited in Stefani & Xenos, 2007, p.110). From the perspective of achieving e-commerce system quality, defining requirements should conform to ISO 9126. Defining requirements involve defining functionality, usability, efficiency, and reliability requirements. Figure 27 demonstrates the activity analysis of defining B2C eService requirements. The definition of requirements needs to include what the service must deliver to the customers, as well as what the management of the service must perform in order to ensure the continued delivery of quality service. What is more, defining these requirements needs to be evaluated against satisfying service quality in general and usability requirements in particular. A further deconstruction of Fig. 27 is done in order to demonstrate further details in terms of usability.

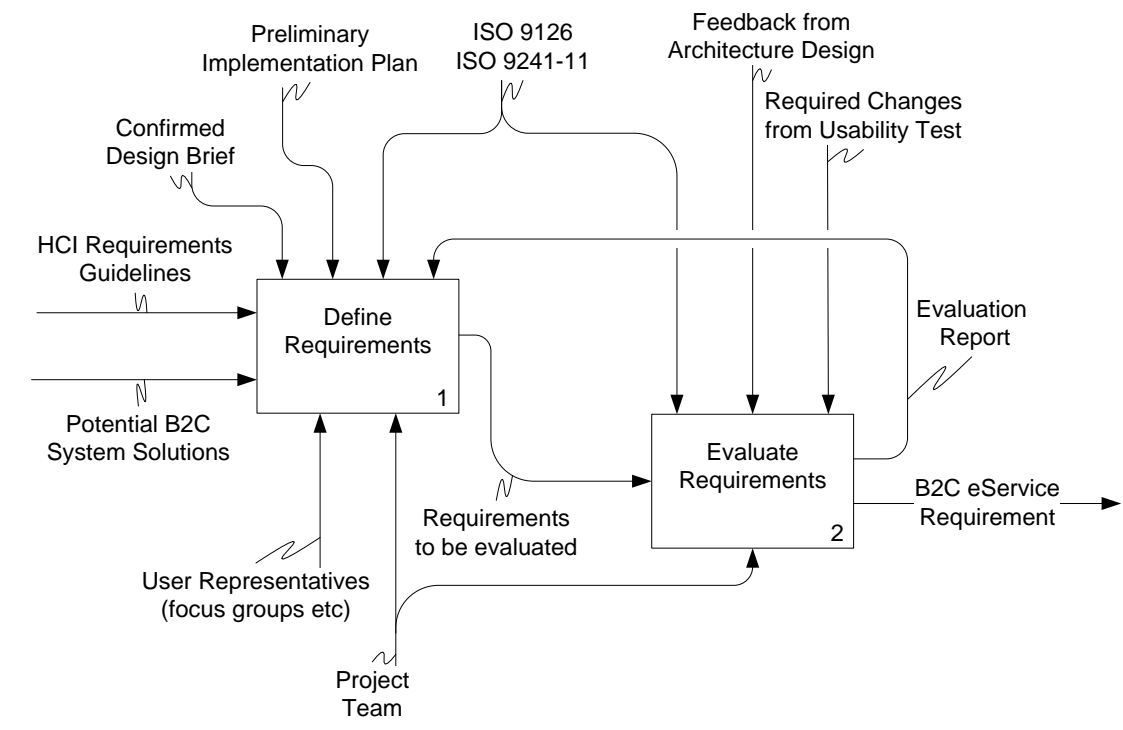

Fig. 27: (P-A12) Define B2C eService Requirements

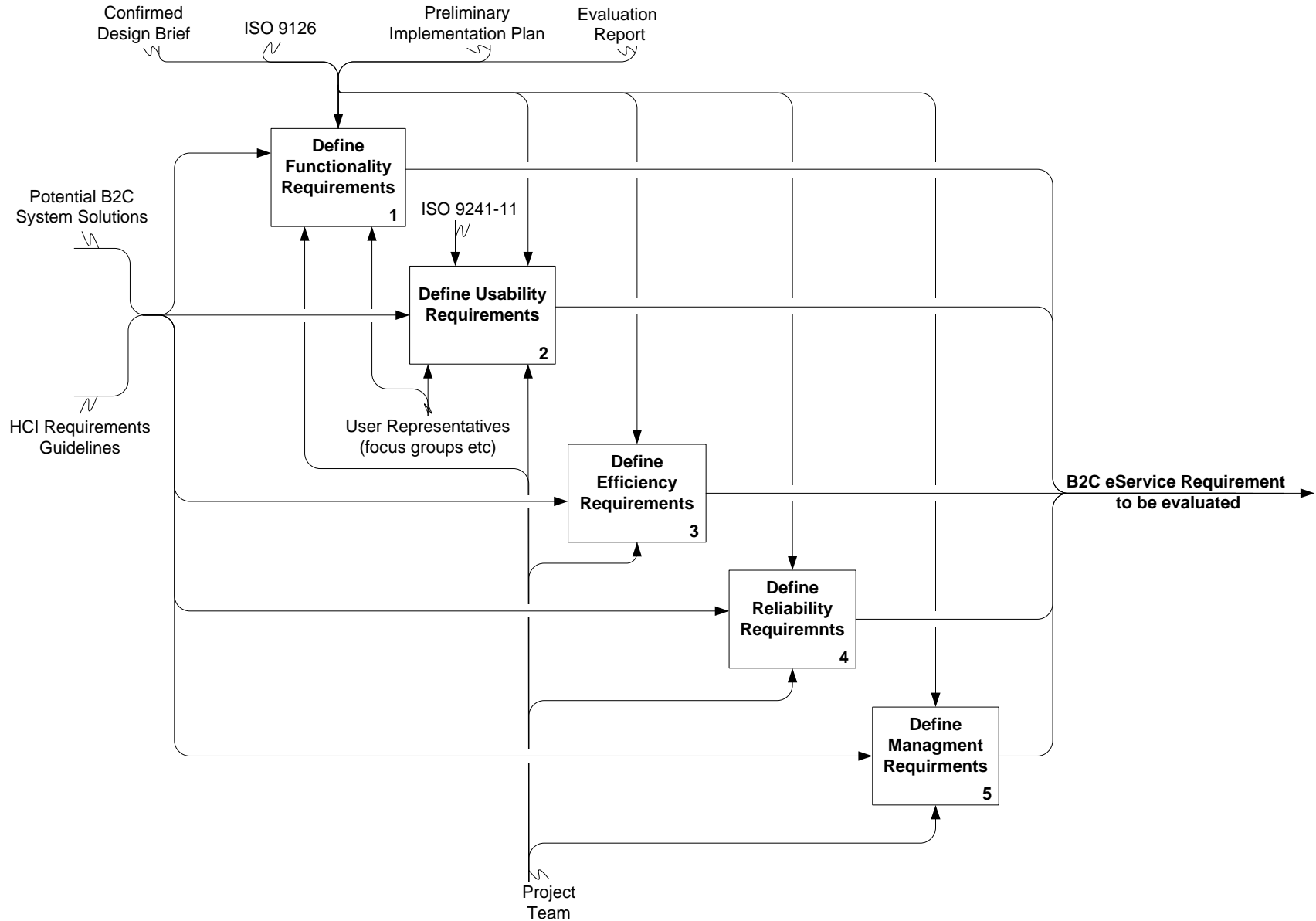

Fig. 28: (P-A121) B2C eService Requirements to be evaluated

Figure 28 demonstrates this further deconstruction of requirements definitions. As discussed in this earlier section, requirements definitions conform to ISO 9126. Thus, defining requirements involves defining functionality, usability, efficiency, and reliability requirements. Furthermore, the definition of requirements needs to include what the management of the service must perform in order to ensure the continued delivery of quality service.

### Functionality Requirements

Functionality requirements refer to functions that are needed in an e-commerce system in order to ensure that the online customers' needs are satisfied. Functionality requirements can be grouped into four characteristics: suitability, accuracy, interoperability and security (Stefani & Xenos, 2007).

Suitability of an e-commerce system involves functions that support navigation, language interpretation and personalization (Stefani & Xenos, 2007). Navigation and language interpretation play important roles in determining ease of use; therefore, they are further discussed in section 4.3.2.3. Personalization needs to be taken into consideration as part of e-commerce system suitability and is a determinant (CSF) of e-commerce success, as discussed in section 2.2.

Accuracy of an e-commerce system refers to online information describing products. Under this term, requirements can be categorized into purchase and search requirements (Stefani & Xenos, 2007). Purchase procedures involve several functions and use metaphors such as shopping carts, checking out, price bargaining (in some cases), account balance (including shipping and postage guarantee), and end with a payment process (Singh, 2002). Searching procedures refer to the search function and is further discussed in the detailed design section due to its involvement with the ease of use (see section 4.3.2.3).

Interoperability of the system refers to the various technologies that are used by an e-commerce system and its environment, and how these technologies interact with one another, such as an e-commerce system interaction with a bank system or other systems to complete the payment process (Stefani & Xenos, 2007).

Security of an e-commerce system refers to the ability of providing a secure transaction environment by the system (Singh, 2002). "*One could further explore the role of security using the ISO/IEC 13335 standard (ISO/IEC 13335-1, 2004) which emphasizes on* [sic] *security techniques, management of information and communications technology security*" (Stefani & Xenos, 2007, p.117)

### *Usability Requirements*

According to ISO 9126, to make a system usable, it should contain the four quality sub-components of: attractiveness, learnability, understandability, and operability. These four characteristics are discussed below.

The interface of an e-commerce system needs to please its users. E-commerce interfaces which offer multimedia technologies, user oriented presentation of products and services, visibility of images etc. attract its customers (Stefani & Xenos, 2007). An unappealing website discourages users from continuing to navigate such a site (Cao et al., 2005). Chen (2001, cited in Cao et al., 2005, p.652) found that "*playfulness is an influential factor*" in attracting users. Playfulness includes tools and designs that attract users' attention with enjoyable constructs (Cao et al., 2005).

Having a learnable e-commerce system, the system's interface should be adapted to the diversity of the customers' familiarity with web interface in general, and online shopping in particular. Showing a website map or informing the users where they are also makes it easy for them to learn. As Stefani & Xenos inform, "*[e]-commerce systems that provide links to purchasing options on the home page; return policy; shipping and delivery information ensure the learnability of navigation and purchase feature*" (2007, p.118). Return users may be offered advanced functions which are hidden from first usage (Shneiderman, 1993).

Online customers need to understand what they are doing before completing their transactions. Having instructive features to ensure understandability by customers improves the chance that they will complete a purchase online. An e-commerce system should have informative characteristics, such as help features, language options, and feedback. The help feature is one component that makes online customers understand how to perform different tasks in the system. This help can be provided through putting up Frequently Asked Questions (FAQs), using interactive assistance functions, or providing contact to a support team by e-mail or phone (Singh, 2002). Language options are another understandability feature, especially when the main target of an e-commerce company is global. However, "*going global means serving international user needs; these needs are shaped by economic, social and cultural factors*" (Stefani & Xenos, 2007, p.118).

In terms of a usable e-commerce system, operability refers to the ability of keeping a system in a functioning and operating condition, including products and business processes that work together to fulfill targeted tasks. An e-commerce system involves several characteristics, such as search features, technical support and help, and personalization (Stefani & Xenos, 2007).

An e-commerce system needs to provide "*searching features that support various forms of keywords make use of logical operators and provide similarity search options*" which add "*advantages for the end-user*" (Stefani & Xenos, 2007, p.119). Search results should be demonstrated and organized according to specific categorizations. Additional services, such as multiple language support, technical support and help, are often overlooked in current implementations. Personalization is another service which is identified in chapter 2 as a critical success factor (see Fig. 2). Stefani & Xenos comment that "*one of the basic prerequisites of the personal store's success is the existence of a complete personal profile. Personalization is the key for a successful recommender system adapted*" to customers preferences (2007, p.119). Recently, many e-Commerce systems have added recommender mechanisms to their applications allow online customers to leave their feedback and recommendations (about the company services and purchased products, e.g. eBay) which clarifies for new customers what previous customers say about the e-commerce company (services, products etc). Hence, these recommendation mechanisms enhance the system usability and trust.

Roben (1998) states that, besides general usability standards, an e-commerce system does have its own set of principles regarding usability. Under this concept, there are significant questions that influence e-commerce system design decisions. For instance, who are the customers? Are they current or new customers, local or global etc.?

In addition, ISO 9241-11 specifies usability requirements through the context of use, and measures usability by three measures: effectiveness, efficiency, and satisfaction. Measuring usability (or usability test) is further discussed in section 4.3.2.2. The context of use comprises four aspects: user, task, equipment, and environment. Thus, specifying usability requirements can be done through the specification of users, tasks, equipment, and environments (ISO 9241-11, 1998).

The specification of the users of B2C eService systems should recognize the remarkable diversity of the users' abilities in using the system. Users can be differentiated into three groups: novice or first time, knowledgeable intermittent, and expert frequent users. Thus, knowing the characteristics of these three groups of users needs to be translated into the system design, making it easy to use for these three classes (Shneiderman, 1993).

The specification of tasks must be determined before the design can proceed (Shneiderman, 1993). All tasks need to be included in this level, "*no matter whether the task is human or automated*", and can be listed as "*activities with inputs, outputs, and controls*" (Bernus, 2003a, p.17).

### Efficiency Requirements

According to ISO/IEC (2001), efficiency is defined as the ability of the system to allow for appropriate performance, relative to the amount of used resources, under declared terms. It refers to a state where system functions are both usable and successful. E-commerce system efficiency refers to navigation, search, and download speed (Santos, 2003).

### Reliability Requirements

Reliability of the e-commerce system means the ability to perform services accurately within the promised time and to reply to customer enquiries, the frequency of updating information, and the accuracy of online transactions (Santos, 2003). It includes three quality sub-characteristics: maturity in functions and services, fault tolerance in interaction, and recoverability in end-users' feedback (Stefani & Xenos, 2007).

***Management Requirements***

The B2C eService also needs its own management, e.g. operational management and control (such as day-to-day monitoring, problem resolutions, maintenance etc.), and tactical management of service (such as monitoring and predicting service levels, planning upgrades, etc.).

***Evaluate against Usability Requirements***

The usability requirements evaluation is a part of other evaluations that need be done for B2C eService requirements. However, the focus of this section is on the usability, and that is why the usability requirements are outlined in Fig. 29 and 'X' and 'Y' refer to other evaluations not discussed here. The B2C eService requirements need to be evaluated in terms of usability requirements to ensure that these requirements are satisfied. The usability evaluation is controlled by the usability standards ISO 9126 and ISO 9241-11, and the evaluation policies and procedures. The evaluation policies and procedures need to be defined by the project manager (as discussed in section 4.3.1). This process is performed by the project team and other participants who need to be allocated by the project management. The output of this process is evaluation reports or required changes.

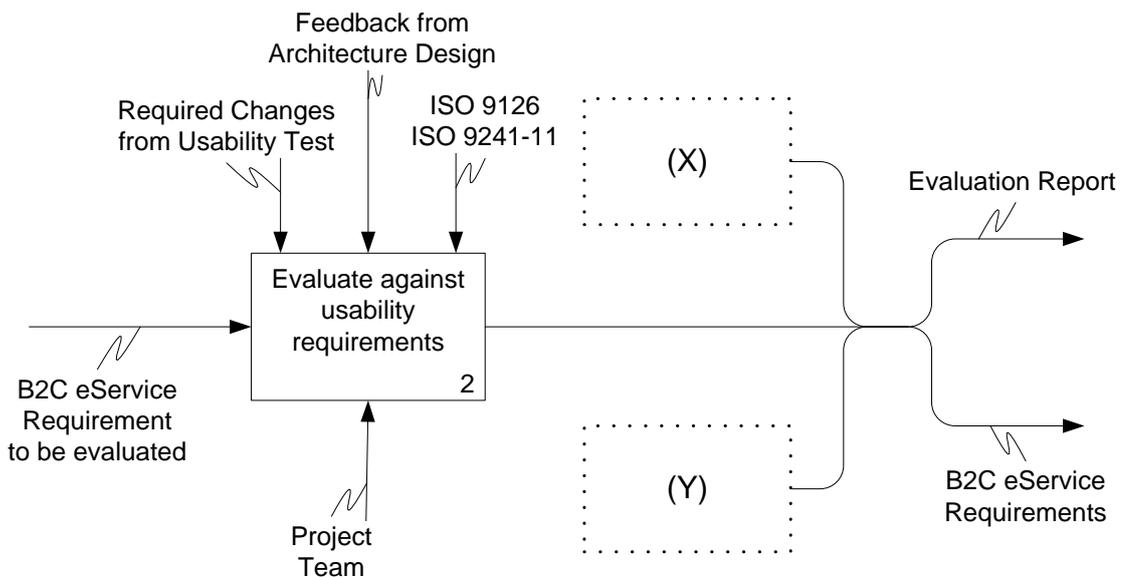

Fig. 29: (P-A122) B2C eService Requirements Evaluation

**Architectural Design**

Architectural design is to define the aggregation of hardware and software elements that make up the B2C system. Architectural design can be further broken down into four phases: divide into subsystems/components (including human and technical components), design human roles, design the user interface, and evaluate the architectural design (see Fig. 30). In addition to technical considerations, the division and the definition of human roles must take into account the HCI Architectural Design Guidelines (to achieve a level of automation that is appropriate for the expected users). Given B2C eService requirements, and a preliminary implementation plan, the project team is able to propose a B2C system master plan after evaluating the proposed architectural design components. The evaluation of architectural design makes changes according to feasibility, quality, risk, cost, and time estimates (Bernus, 2003a). The proposed B2C system master plan needs to be reviewed by HQ, the project manager, and the Business Development team, as discussed in section 4.3.1 (see Fig. 20). This revision is needed for the approval of the proposed master plan. Architectural design may also receive feedback (required changes) from the detailed design phase and usability testing (see section 4.3.2.2 and Fig. 32).

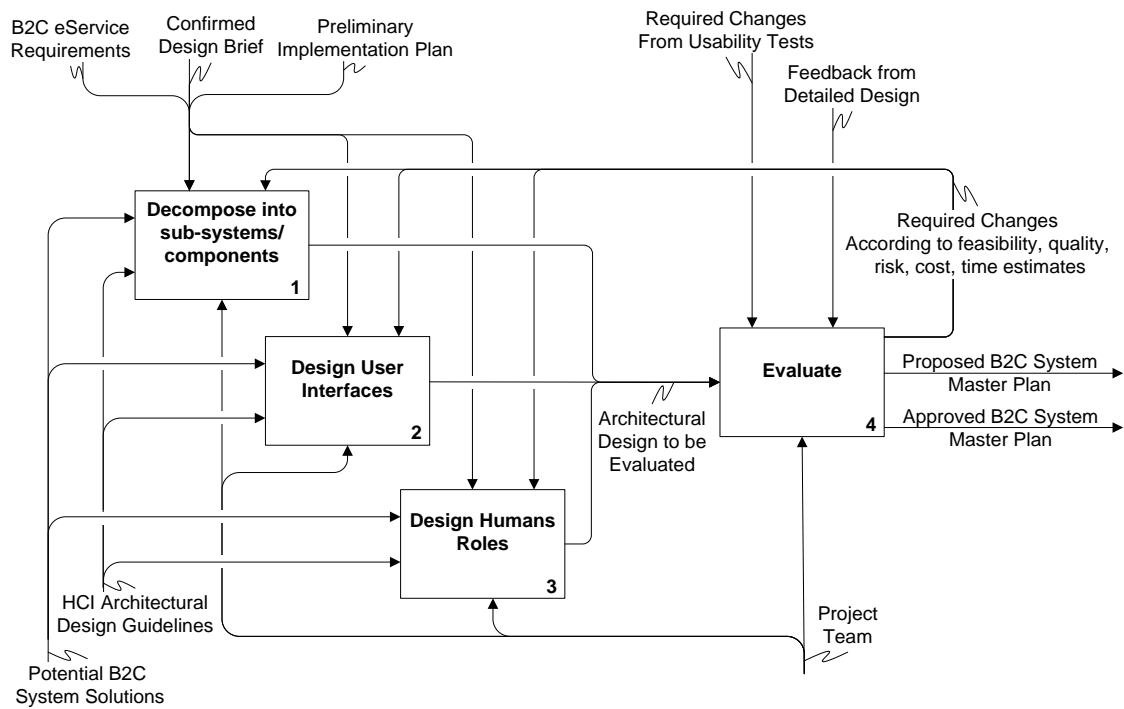

Fig. 30: (P-A13) B2C eService architectural Design

*User Interface Design*

E-commerce websites must accommodate almost all customers - novice, beginners and experts (Najjar, 2001); therefore, special care must be taken for designing the page format, catalog, navigation, personalization, registration, and checkout.

Kim and Lee (2002) divided design factors into two independent perspectives: transaction process and web document architecture. Furthermore, these two perspectives are further structured into phases (see Fig. 31).

 (1) The transaction process includes four phases: information, agreement, settlement, and environment. The information phase includes design factors that mostly relate to the process of searching for specific products or browsing the system for interesting items. The agreement phase includes design factors that relate to order and negotiation facilities. Settlement phase starts after customers have completed the processes in the agreement phase, and includes design factors that relate to the payment process, delivery of products ordered, and after sales support (interaction, guarantee, claim, refund etc.). Environment phase includes design factors that relate to data transmission (speed, security, privacy, policy, and standard contract forms etc.).

 (2) The web document architecture describes the connections between four elements: content, structure, interaction, and presentation. The content element relates to the decision of which type of information is relevant to the description of products or services. The structure element relates to categorizing products and features according to relevance, similarity etc. The interaction element is to enable customers to interact effectively with the system, including browsing and searching facilities. The presentation element relates to the arrangement and design of the screen, including various items: pixels, sizes, backgrounds,

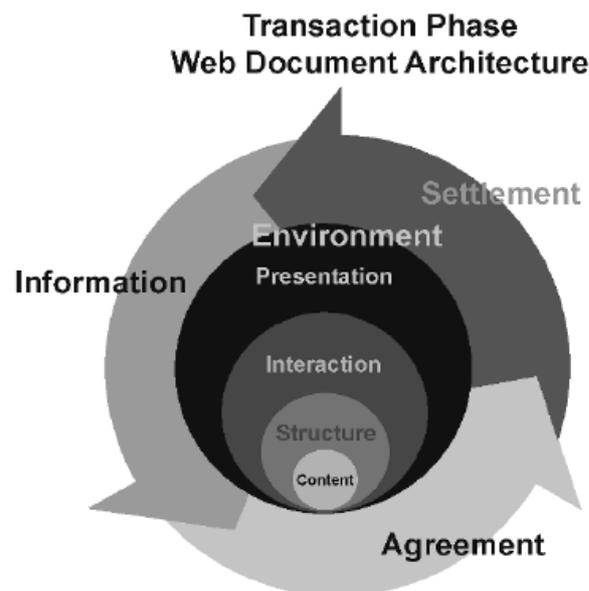

Fig. 31: E-Commerce System Design Factors
Source: Kim and Lee, 2002, p.189

navigation buttons, logos, images and icons, colours, polygons, and pages layouts.

All of the above need to take into account the HCI Architectural Design guidelines determined in previous phases.

### *Designing Human Roles*

Human roles involve activities and tasks that are performed by the users of the system (including customers, Customer Relationship Management (CRM) team, sales team etc.). Designers have to determine what the tasks that prospective users have to perform when using the systems. These tasks will be determined in usability user interface and by expected skills and background knowledge of the prospective users. The level of the tasks that are expected to be performed by a normal customer (e.g. a customer purchasing a bike) is not the same as the level of the tasks that are expected to be performed by a professional user (e.g. a mechanic purchasing motorbike parts). Hence, it is necessary to consider human diversity (in abilities, skills, knowledge, cultures etc.) when designing the system (Shneiderman, 1993). Then, the process of designing human roles requires taking into consideration the Human-Computer Interaction (HCI) guidelines (as per in A3 – Fig. 30) which include reference to methods that can be used for this purpose.

### 4.3.2.2 Implementing the eService System

Implementing the eService system requires an approved B2C system master plan (which is an outcome of the architecture design in 'Initiating the eService system' – see Fig. 25), a final implementation plan (GANTT chart, deadlines, budget, etc. which come from the management level – see Fig. 21), as well as having to take into account standards as noted (ISO 9126 and ISO 9241-11) to guide the project participants (Project Team, the Business Unit team, and usability experts) whose common task is to deliver the eService system.

The eService system implementation process is further divided into four sub-activities: detailed design, release to operation, usability test, and training in usability. The first two sub-activities (detailed design and release to operation) are based on the business model (the fifth and sixth life cycle phases of the B2C eService entity); whereas usability testing is the operation for testing purposes. The training of project members is included here to show that usability experts may have to be trained if they are not available (however, logically, this activity is really part of project building).

The process of the eService system implementation starts with the detailed design (which requires the observation of detailed design guidelines). Given the approved B2C system master plan, relevant standards (ISO 9126 and ISO 9241-11) and the final implementation plan, the project team (with user representatives' participation) is able to work out a detailed system design (note that according to the life cycle model used, the detailed design of the software *includes* coding). Feedback (as a control) for the architectural design is another outcome of this process (if the detailed design encounters problems that only the architectural design can solve).

Given the designed system, the process participants (project team, the Business Unit team and external services providers) are able to build the system (release it to operation) to be ready for operation (or for testing the operation). The first the system is released to test the operation. These tests are to ensure that the provided service meets usability requirements. The usability tests are guided by the approved B2C system master plan and the relevant standards (ISO 9126 and ISO 9241-11). Given these guidelines and usability test guidelines (approved as part of Adopted HCI guidelines), the usability experts, user participants (focus group etc) and project team are able to perform this process. In case the project team is not ready for this process, another activity is needed (training in usability, which can be performed by usability experts) to enable the project team to participate in the usability testing process. The feedback (usability test report) of the usability tests is feedback on several previous phases (in the eService system development process, see Fig. 25) which is used as a control to perform the required changes.

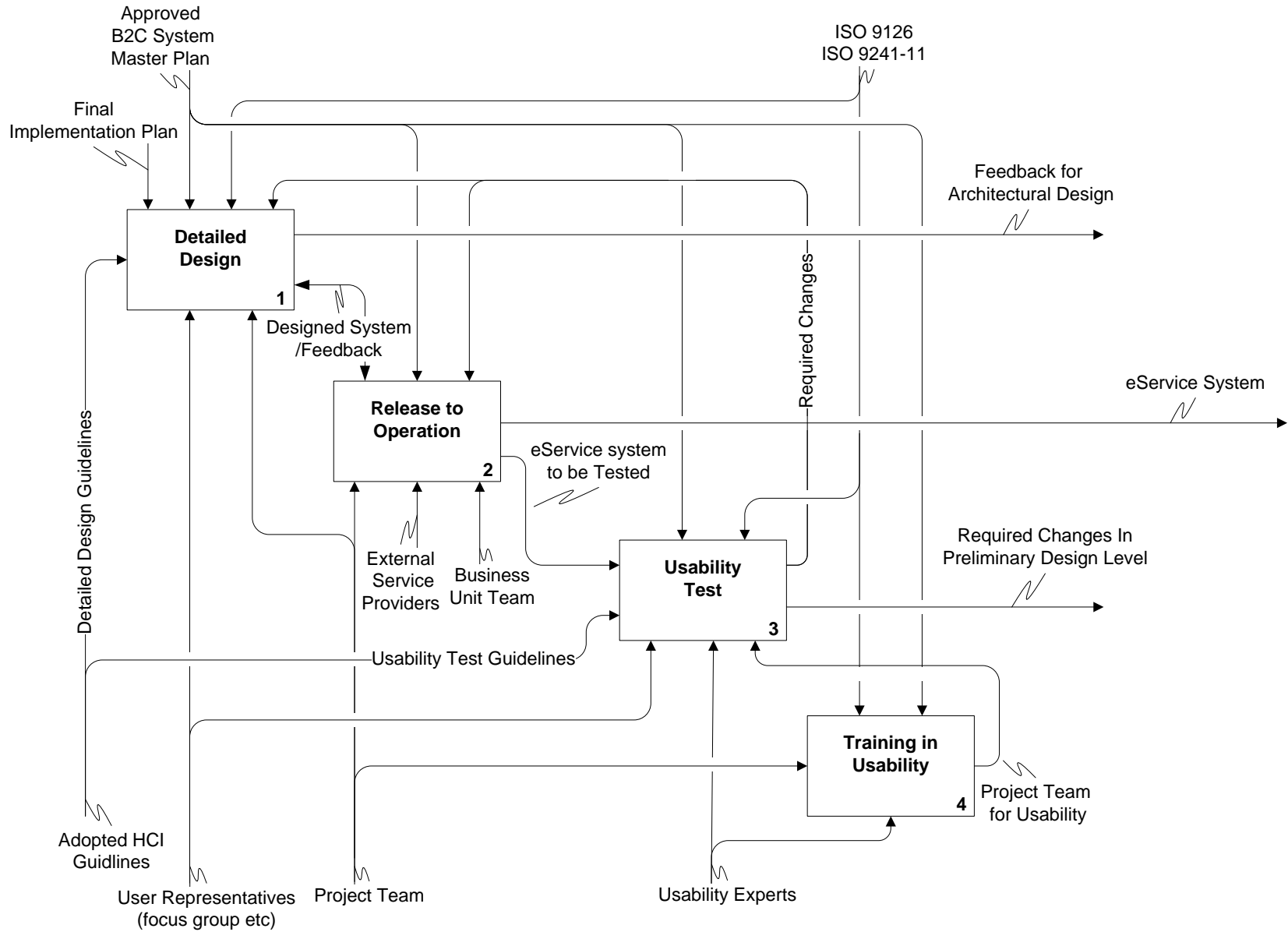

Fig. 32: (P-A2) Implementing the eService System

**Detailed design**

The detailed design activity includes several activities (detailed design of user interface, data services/data bases, eService applications etc). Since 'ease of use' is being analysed, the focus is on the user interface, because it is the connection between online customers and the B2C e-Commerce system. The detailed design process of the user interfaces are considered to be important in providing an easy to use service. The description of the detailed interface design can be divided into three groups: Page design, Navigation and Checkout (Najjar, 2001). Page design includes various aspects of web page design, such as page layout, text appearance, graphics usage, colours, and font size. The Navigation group includes search facilities, order and negotiation of services, and interaction design. The Checkout group includes registration, payment process, shipping alternatives, privacy and security policies etc.

*Page design*

A homepage should communicate to online customers what the website is about and what services it provides. All major options must be demonstrated on the homepage. A homepage needs to be accessed easily from other pages. In case there are major changes to be done on a homepage, return users need to be alerted to avoid the disadvantage of surprise (Leavitt & Shneiderman, 2006).

Information needs to be clearly headed. To enable users to adopt systematic page scanning, information needs to be structured according their importance level and relation (Santos, 2003). Each content page needs to be structured and organized to facilitate scanning (Leavitt & Shneiderman, 2006). Hierarchical structure facilitates easy understanding and focusing attention (Kim & Lee, 2002). Background colours can be used for easy understanding of the groupings of related information. What is more, pages need to be not too crowded with information and avoid long sentences. Instead, the design should use appropriate graphics, tables and visualization techniques which are more understandable (Cao et al., 2005). Graphics and images should convey intended messages (Silver, 2005) and be small in size, yet be enlarged if needed (Udo & Marquis, 2001). Critical information and data should be highlighted. The page title should be descriptive of what the page contains. Font size and colours must be familiar to users; for example, the font size should be at least 12-points and colours should not be disturbing to the eyes (Leavitt & Shneiderman, 2006).

There are also other page format guidelines which need to be taken into consideration. The need to quickly download web pages may require the designer to limit the usage of graphics and Flash animations. Background images should be simple (Najjar, 2001). Horizontal scrolling is annoying for many users, so the page design needs to be fitted horizontally. Pages need to be set at an appropriate length. The labels of the links should be meaningful and understandable. To indicate visited links, colour changes should be used. The length of links needs to be appropriate. In case of a link leading to an external page, it is good practice to indicate this (Leavitt & Shneiderman, 2006).

*Navigation*

In addition to links, the second most important means of navigation is the search function. Therefore, to improve the effectiveness of the search function, it needs to include "*meta tag tools, thesauruses, alternate spelling, and database search engines*" (Najjar, 2001, p.3). The option to perform a search or to refine the search needs to be available on any page and should have suitable alternatives for the search, such as search by product names, categories, items numbers, and prices (Leavitt & Shneiderman, 2006; Kim & Lee 2002). In the first results page, the most relevant hits should appear (Kim & Lee, 2002) and specific hints to improve search performance should be included (Leavitt & Shneiderman, 2006).

Navigation options should be in the same location on each page (Udo & Marquis, 2001). It is important to provide feedback on the users' location so that they know where they are on the site (Silver, 2005). The number of major sections should be limited, and their names must be familiar to avoid confusing users. The site should allow direct entry into the product catalogue, and the user should have effective, easy and different tools to sort products by price, brand, colour etc., select and compare products side by side, have access to each product's details page, add or remove products from a list, and reduce the number of options. Ordering and negotiation facilities take important place in the navigation. Thus, providing interaction mechanisms such as e-mail and chat enables customers to interact with the site (Kim & Lee, 2002). Furthermore, Frequently Asked Questions (FAQs) and a help section should be clearly seen on each page to help customers, especially novice users, understand different tasks (Leavitt & Shneiderman, 2006).

*Checkout*

The Checkout stage starts after a customer decides to complete a purchase order. Najjar (2001) proposes to not require registration before checking out. It is much better to allow new users to enter their shipping and billing details and to then make registration optional, as Amazon.com does. For effective and easy registration, it is required to enter an e-mail address and a password, as eBay does, and to limit repeating the entry of the same information by using ticking options. For increasing trust, it is essential to provide obvious links to privacy and security policies and to certify the site's security and information privacy by consumer groups (e.g. TRUSTEe or BBBOnLine), and to show their logos on the checking out page (Najjar, 2001).

Before requiring a customer to pay, it is important to show a complete and editable order summary that includes the product name, price, shipping cost and additional fees. The site should clearly notify the user when there is an error by using an obvious error message at the top in red, and should highlight the fields or text where the error has occurred so that it can be easily fixed (Najjar, 2001). Providing several reliable payment methods, such as credit card, EFT, electronic cheques, smart cards, e-cash and intermediaries such as PayPal and iTransact, gives customers flexibility to pay according to their preference (Pearlson & Saunders, 2006). Singh (2002) informs that providing e-payment methods (such as PayPal) is more practical in e-commerce to facilitate the payment process for customers. Agrawal et al. (2001, cited in Najjar, 2001) strongly recommends providing clear links to enhance customer protection features, e.g. security protection and delivery guarantees, and order confirmation including all important information; namely, the shipping address, total cost, order number, cancellation instructions, contacts details etc.

**Usability Tests**

The usability tests performed by usability experts, user representatives (focus group etc.) and the project team are very important in order to ensure the quality of the system. The feedback (usability test report) of this process may be sent back to several previous phases (in the eService system development process) as a control to perform the changes required by the usability test results.

Leavitt and Shneiderman (2006) state that two major considerations need to be taken into account when carrying out a usability test. Firstly, it is to guarantee using the best possible testing method. In general, the best way is to carry out a test where user representatives interact with representative scenarios while testers collect testing data (to be provided to designers in order to develop, improve and maintain the system). Secondly, it is to use an iterative approach. A system needs to be tested again after making changes in order to get accurate results.

Selecting an appropriate number of usability test participants plays an important part in determining the usability testing success. Moreover, an e-commerce system needs to be tested by usability experts and representative users. Usability experts inspect and evaluate the system and representative users test the system performance. Furthermore, it is important to select test users from different expertise levels (experts, beginners and novices) (Zou, Zhang & Zhao, 2007).

Rich and McGee (2004, cited in Leavitt & Shneiderman, 2006, p.191) make the following recommendations:

- Prioritizing tasks when deciding which usability issues to fix first.
- User expectations of the difficulty of each task can be assessed using the Usability Magnitude Estimation (UME) method.
- Participants will judge the difficulty or ease of a task twice; once before trying to do it, and another after trying to complete it.
- The prioritization can be categorized according to the following four ratings:
  - *"Tasks that were expected to be easy, but were actually difficult;*
  - *Tasks that were expected to be difficult, but were actually easy;*
  - *Tasks that were expected to be easy and were actually easy; and*
  - *Tasks that were expected to be difficult and were difficult to complete."*

# 5. Conclusions and Future work

The literature identifies a wide range of issues which can be considered as the CSFs for B2C e-Commerce systems. From the literature, a list was compiled of 29 CSFs of B2C e-Commerce systems. The identified CSFs were systematized in a new structure in order to make it easy to concentrate on underlying CSFs. Using a generic business model, the scope of these CSFs was developed; and to understand the process of the development project, the process was modeled using an activity modeling language to demonstrate what needs to be done in order to ensure that the CSFs are satisfied. The result of this thesis is that in order to ensure that CSFs are satisfied, one must consider:

- *The quality of the B2C system itself.*

- *The quality of the eService development project.*

- *The quality of the B2C system's management process.*

There are a number of limitations in this study that must be acknowledged. First, the validity of the results hinges on the correct interpretation of the CSFs reported in the literature; future research needs to be done after this thesis is complete to confirm this interpretation. Second, the limited scope of this thesis (time and resources) does not allow the analysis of all CSFs (only one example CSF was chosen, namely 'ease of use') to show what needs to be done (by an enterprise) in order to satisfy them.

Furthermore, this research could be extended using other methods that can add value to this study. One possibility is to do a case study with observation of the use of these results, which would provide feedback to the researcher. Another method is to conduct action research, which allows the researcher to practice in the development of a B2C system. This method of conducting research could add value and improve on the results instead of only confirming the validity/usability of the present research outcomes.

Moreover, the business model used in this study can be enriched by mapping other CSFs onto it in a similar manner by

- adding the relevant information (such as standards) to the developed life cycle process model, or

- adding the necessary new activities to the process model

to satisfy these CSFs.

# References

**[1]** Aberg, F & Shahmehri, N 2000, 'The role of human Web assistants in e-Commerce: an analysis and a usability study', *Internet Research: Electronic Networking Applications and Policies*, vol. 10, no. 2, pp. 114-25.

**[2]** Abran, A, Khelifi, A, Suryn, W & Seffah, A 2003, 'Usability Meanings and Interpretations in ISO Standards', *Software Quality Journal*, vol. 11, no. 4, pp. 325-38.

**[3]** Atchariyachanvanich, K, Okada, H & Sonehara, N 2007, 'What Keeps Online Customers Repurchasing through the Internet?' ACM SIGecom Exchanges, vol. 6, no. 2, pp. 47-57.

**[4]** Bernus, P 2003, *6118CIT Introduction to Enterprise Architecture: course notes*, Griffith University, Brisbane, viewed 5 May 2008, <http://www.cit.gu.edu.au/~s362104/ea/S1.2008/index.html>.

**[5]** Bernus, P 2003b, 'Modelling Function and Information', in P Bernus, L Nemes & G Schmidt (eds), *Handbook on Enterprise Architecture*, Springer, pp. 417-33.

**[6]** Chen, C & Yang, S 2006, 'E-Commerce and Mobile Commerce Applications Adoptions', in M Khosrow-Pour (ed.), Encyclopedia of E-Commerce, E-Government, and Mobile Commerce, Idea Group Reference, USA, vol. 1, pp. 284-90.

**[7]** Chen, H, Chen, Q & Kazman, R 2007, 'The Affective and Cognitive Impacts of Perceived Touch on Online Customers' Intention to Return in the Web-based eCRM Environment', *Journal of Electronic Commerce in Organisations*, vol. 5, no. 1, pp. 69-91.

**[8]** Chen, Z, Li, H, Kong, S, Hong, J & Xu, Q 2006, 'Process Simulation for E-Commerce Systems', in M Khosrow-Pour (ed.), *Encyclopedia of E-Commerce, E-Government, and Mobile Commerce*, Idea Group Reference, USA, vol. 1, pp. 934-40.

**[9]** Dubelaar, C, Sohal, A & Savic, V 2005, 'Benefits, impediments and critical success factors in B2C E-business adoption', *Technovation*, vol. 25, pp. 1251-62.

**[10]** Gefen, D & Straub, D 2004, 'Consumer trust in B2C e-Commerce and the importance of social presence: experiments in e-Products and e-Services', *The International Journal of Management Science*, vol. 32, pp. 407 – 24.

**[11]** Green, D & Pearson, M 2006, 'Development of a web site usability instrument based on ISO 9241-11', *Journal of Computer Information System*, vol. 47, pp. 66-72.

**[12]** Hassanein, K & Head, M 2004, 'Building Online Trust through Socially Rich Web Interfaces', paper presented to Proceedings of the 2nd Annual Conference on Privacy, Security and Trust, Fredericton, New Brunswick, Canada, 15-22.

**[13]** Holsapple, C & Sasidharan, S 2005, 'The dynamics of trust in B2C e-commerce: a research model and agenda', *Information Systems and E-Business Management*, vol. 3, pp. 377-403.

**[14]** IFIP-IFAC Task Force 2003, 'GERAM: The generalized enterprise reference architecture and methodology, IFIP-IFAC Task Force on Architectures for Enterprise Integration', in P Bernus, L Nemes & G Schmidt (eds), *Handbook on Enterprise Architecture*, Springer, pp. 417-33.

**[15]** ISO 1998, 'ISO 9241-11: Ergonomic requirements for office work with visual display terminals (VDTs) - Guidance on usability', International Organization for Standardization.

**[16]** KBSI 2006, *IDEFØ Function Modeling Method* Knowledge Based Systems Inc, viewed 8 May 2008, <http://www.idef.com/idef0.html>.

**[17]** Kim, J & Lee, J 2002, 'Critical design factors for successful e-commerce systems', *Behaviour & Information Technology*, vol. 21, no. 3, pp. 185-99.

**[18]** Leavitt, M & Shneiderman, B 2006, Research-Based Web Design & Usability Guidelines, U.S. Government Printing Office.

**[19]** Lepouras, G & Vassilakis, C 2006, 'Adaptive Virtual Reality Shopping Malls ', in M Khosrow-Pour (ed.), *Encyclopedia of E-Commerce, E-Government, and Mobile Commerce*, Idea Group Reference, USA, vol. 1, pp. 1-6.

**[20]** Molla, A & Licker, PS 2001, 'E-commerce Systems Success: an attempt to extend and Respecify the Delone and Maclean Model of IS success', *Journal of Electronic Commerce Research*, vol. 2, no. 4, pp. 131-41.

**[21]** Najjar, L 2001, 'E-commerce User Interface Design for the Web', paper presented to Human-Computer Interaction International Conference, Mahwah, NJ, USA.

**[22]** NIST 1993, *Announcing the Standard for Integration Definition For Function Modeling (IDEF0)*, National Institute of Standards and Technology.

**[23]** Pearlson, K & Saunders, C 2006, 'Doing Business on The Internet', in *Managing and Using Information System*, John Wiley and Sons, 161-191, New Caledonia.

**[24]** Ribbink, D, vanRiel, A, Liljander, V & Streukens, S 2004, 'Comfort your online customer: quality, trust and loyalty on the Internet', *Managing Service Quality*, vol. 14, pp. 446-56.

**[25]** Roben, JA 1998, 'Creating Usable E-Commerce Sites', *StandardView*, vol. 6, no. 3, pp. 110-6.

**[26]** Ryan, G & Valverde, M 2006, 'Waiting in line for online service: a qualitative study of the user's perspective', *Information System Journal*, vol. 16, pp. 181-211.

**[27]** Santos, J 2003, 'E-service quality: A model of virtual service quality dimensions', *Managing Service Quality*, vol. 13, no. 3, pp. 232-46.

**[28]** Schodeler, D & Madeja, N 2004, 'Is Customer Relationship Management a Success Factor in Electronic Commerce?' *Journal of Electronic Commerce Research*, vol. 5, no. 1, pp. 38-53.

**[29]** Shankar, V, Smith, A & Rangaswamy, A 2003, 'Customer satisfaction and loyalty in online and offline environments', *International Journal of Research in Marketing*, vol. 20, pp. 153–75.

**[30]** Shneiderman, B 1993, *Designing the User Interface: Strategies for Effective Human-Computer Interaction*, Addison Wesley.

**[31]** Silver, M 2005, *Exploring Interface Design: Proven Techniques for Creating Compelling & Usable Interfaces for Multimedia & the Web*, Thomson Delimar Learning, New York.

**[32]** Singh, M 2002, 'E-services and their role in B2C e-commerce', *Managing Service Quality*, vol. 12, no. 6, pp. 434-46.

**[33]** Song, J & Zahedi, F 2001, 'Web design in e-commerce: a theory and empirical analysis', paper presented to Proceedings of the 22nd International Conference on Information Systems, New Orleans, Louisiana

**[34]** Stefani, A & Xenos, M 2007, 'E-commerce system quality assessment using a model based on ISO 9126 and Belief Networks', *Software Qual J*, vol. 16, pp. 107–29.

**[35]** Sung, T 2006, 'E-commerce Critical Success Factors: East vs. West ', *Technology Forecasting and Social Change*, vol. 73, pp. 1161-77.

**[36]** Terry, J & Standing, C 2004, 'The Value of User Participation in E-Commerce System Development', *Information Science Journal*, vol. 7, pp. 31-44.

**[37]** Udo, G & Marquis, G 2001, 'Factors affecting e-commerce Web site effectiveness', *The Journal of Computer Information Systems*, vol. 42, no. 2, pp. 10-6.

**[38]** Williams, TJ, Rathwell, GA & Li, H 2001, *A Handbook on Master Planning and Implementation for Enterprise Integration Programs*, West Lafayette, Indiana.

**[39]** Wirtz, B & Lihotzky, N 2003, 'Customer Retention Management in the B2C Electronic Business', *Long Range Planning Journal*, vol. 36, pp. 517-32.

**[40]** Yeo, A & Chiam, M 2006, 'E-Customer Loyalty', in M Khosrow-Pour (ed.), *Encyclopedia of E-Commerce, E-Government, and Mobile Commerce*, Idea Group Reference, USA, vol. 1, pp. 334-9.

**[41]** Zou, Y, Zhang, Q & Zhao, X 2007, 'Improving the Usability of e-Commerce Applications Using Business Processes', *IEEE Transactions on Software Engineering*, vol. 33, no. 12, pp. 837-55.

## Bibliography

Corritore, C, Marble, R, Wiedenbeck, S, Kracher, B & Chandran, A 2005, 'Measuring Online Trust of Websites: Credibility, Perceived Ease of Use, and Risk', paper presented to Eleventh Americas Conference on Information Systems, Omaha, NE, USA.

Doculabs 2003, *Planning and Building an Architecture that Lasts: The Dynamic Enterprise Reference Architecture, Doculabs Marketfocus Report*.

Ghaoui, C 2006, *Encyclopedia of Human Computer Interaction*, Idea Group Reference, Liverpool.

Giaglis, G, Klein, S & O'Keefe, R 2002, 'The Role of Intermediaries in Electronic Marketplace: developing a contingency model', *Information System Journal*, vol. 12, pp. 231-46.

Goldsborough, R 2002, 'The Influence of Active Online User', *Academic Research Library*, vol. 5, no. 19, pp. 1-30.

Huarng, AS & Christopher, D 2003, 'Planning an effective Internet retail store', *Marketing Intelligence & Planning*, vol. 21, no. 4, pp. 230-8.

Khosrow-Pour, M 2006, *Encyclopedia of E-Commerce, E-Government, and Mobile Commerce*, Idea Group Reference, USA.

Liua, C & Arnett, K 2000, 'Exploring the factors associated with Web site success in the context of electronic commerce', *Information & Management*, vol. 38, pp. 23-33.

Schlaeger, C & Pernul, G 2005, 'Authentication and Authorization Infrastructures in B2C e-commerce', *Springer-verlag Berlin Heidelberg*, pp. 306-15.

Sharma, S, Gupta, J & Wickramasinghe, N 2006, 'A framework for designing the enterprise-wide e-commerce portal for evolving organizations', *Electronic Commerce Res*, vol. 6, pp. 141-54.

Zhou, L, Dai, L & Zhang, D 2007, 'Online Shopping Acceptance Model - A Critical Survey of Consumer Factors in Online Shopping', *Journal of Electronic Commerce Research*, vol. 8, pp. 41-62.